# Concurrent daytime and nighttime heatwaves in the late 21st century over the CORDEX-East Asia phase 2 domain using Multi-GCM and Multi-RCM Chains

**Young-Hyun Kim[1], Joong-Bae Ahn[1*], Myoung-Seok Suh[2], Dong-Hyun Cha[3], Eun-Chul Chang[2], Seung-Ki Min[3], Young-Hwa Byun[5] and Jin-Uk Kim[5]**

[1]Department of Atmospheric Sciences, Pusan National University, Busan, Korea

[2]Department of Atmospheric Science, Kongju National University, Gongju, South Korea

[3]Department of Urban and Environmental Engineering, Ulsan National Institute of Science and Technology, Ulsan, South Korea

[4]Division of Environmental Science and Engineering, Pohang University of Science and Technology, Pohang, South Korea

[5]Climate Change Research Team, National Institute of Meteorological Sciences

*Corresponding Author : Joong-Bae Ahn (E-mail : jbahn@pusan.ac.kr)

**Abstract**

The adverse impacts of extreme heat on human health when a concurrent daytime and nighttime heatwave (CDNHW) occurs are greater than when daytime or nighttime heatwaves occur individually, because of the reduced recovery time from heat exposure. This study projects increases in CDNHW over the whole of East Asia under two Representative Concentration Pathway scenarios (RCP2.6 and RCP8.5) and two Shared Socioeconomic Pathway scenarios (SSP1-2.6, and SSP5-8.5). The daily maximum and minimum temperatures, which are used to define a CDNHW, are calculated of 3-hourly temperatures of 25km horizontal resolution produced by 12 general circulation model and regional climate model chains participating in the Coordinated Regional Climate Downscaling Experiment East Asia phase 2 project. In Historical simulation (1981-2005), occurrence period and occurrence rate of CDNHW from April to September area-averaged in East Asia are 10.9 days and 0.9%, respectively. In projections for the future (2071-2100), occurrence period and occurrence rate of CDNHW will be three weeks and 3.7% (RCP2.6), two month and 20.5% (RCP8.5), two month and 15.6% (SSP1-2.6), and three months and 45.7% (SSP5-8.5). In addition, it is expected that the CDNHW intensity will increase, and the spatial extent of CDNHW will be extended. Although a CDNHW lasting less than three days is the most common, the proportion of CDNHWs lasting more than ten days, compared to the total CDNHW frequency, will increase to 1.2% (RCP2.6), 7.2% (RCP8.5), 6.1% (SSP1-2.6), and 17.3% (SSP5-8.5) from 0.2% (Historical). Both occurrence rate and intensity of CDNHW will increase to a relatively large extent in Indochina, East and West China and India. If the current greenhouse gas emissions continue, East Asia will experience unprecedented heat stress because the frequency and intensity of CDNHWs, which rarely occur during present-day, will increase significantly over all regions by the end of the 21$^{st}$ century.

## 1. Introduction

The consecutive occurrence of hot days and hot nights increases the adverse impacts of extreme heat on human health, because the recovery time from heat exposure is reduced (Fischer and Schar 2010, Opitz-Stapleton *et al* 2016, Wang *et al* 2020a, Zhang *et al* 2020). According to recent studies, the frequency of hot days and hot nights is increasing, and that trend will accelerate in East Asia, including South Korea (Im *et al* 2017, Hong *et al* 2018, Choi and Lee 2019), China (Chen and Li 2017, Li *et al* 2017, Ding *et al* 2018, Guo *et al* 2018, Li *et al* 2018, You *et al* 2018, Chen and Dong 2019, Hu and Sun 2020, Wang *et al* 2020b, Wu *et al* 2020, Zhang *et al* 2020), Mongolia (Tong *et al* 2019), India (Panda *et al* 2014, Mukherjee and Mishra 2018, Ahsan *et al* 2022), Vietnam (Opitz-Stapleton *et al* 2016), most regions of Asia (Dong *et al* 2018), the Northern Hemisphere (Wang *et al* 2020a), and globally (Zhu *et al* 2021).

It is expected that heat stress will be exacerbated due to global warming, and it is especially necessary to analyze future changes in the concurrent daytime and nighttime heatwave (CDNHW), a phenomenon that extreme hot temperatures last from day to night. Many previous studies focused on future changes in hot days or hot nights separately (Opitz-Stapleton et al 2016, Im et al 2017; Guo et al 2018; Li et al 2018; Wang et al 2019; Hu and Sun 2020, Wu et al 2020, Ahsan et al 2022). Although most previous studies projected CDNHWs in specific regions or countries, the threshold for a CDNHW varied from study to study (Mukherjee and Mishra 2018; Ullah *et al* 2022 for India, Su and Dong 2019, Wang et al 2020b for China). In other words, there has never been a comprehensive and integrated study that analyzed CDNHWs over the whole of East Asia. Some studies showing future changes in CDNHWs by applying a united threshold to the northern hemisphere or on a global scale used data from the general circulation model (GCM) (Wang et al 2020a, Zhu et al 2021, Ma et al 2022).However, those studies based on GCM projections have limited data resolution when it comes to simulating extreme events related to heat stress in East Asia with its complex topography.

In this study, we project a concurrent daytime and nighttime heatwave by applying a united threshold that considers spatial and temporal climate characteristics over the whole of East Asia by the end of the 21$^{st}$ century by using 3-hourly temperature data with a 25km horizontal resolution produced by 12 general circulation model–regional climate model (GCM-RCM) chains. The future simulations were based on two Representative Concentration Pathway (RCP) scenarios (RCP2.6 and RCP8.5) for GCMs from coupled model inter-comparison project phase 5 (CMIP5) and two Shared Socioeconomic Pathway (SSP) scenarios (SSP1-2.6 and SSP5-8.5) for GCMs from CMIP phase 6 (CMIP6).

## 2. Data and analysis method

### 2.1. Model and observation data sets

This study utilizes the 3-hourly temperature data with a 25km horizontal resolution from 12 GCM-RCM chains in the Coordinated Regional Climate Downscaling Experiment East Asia phase 2 (CORDEX-EA2) project (Figure 1). Eight GCM-RCM chains (HG2_CCLM, HG2_RegCM, HG2_MM5, MPI_WRF, MPI_CCLM, MPI_MM5, GFDL_WRF, and GFDL_RegCM) are projections of three GCMs from CMIP5 under RCP2.6 and RCP8.5 scenarios that are dynamically downscaled using three or two RCMs. The other four GCM-RCM chains (UKESM_WRF, UKESM_CCLM, UKESM_RegCM, and UKESM_GRIM) are projections of one GCM from CMIP6 under SSP1-2.6 and SSP5-8.5 scenarios dynamically downscaled using four RCMs. The configurations of the 12 GCM-RCM chains are summarized in table S1. The multi-model ensemble (ENS), which is the average of GCM-RCM chains with equal weighting, is mainly used to enhance the reliability of future climate projections compared to a single model. The averages for HG2_CCLM, HG2_RegCM, HG2_MM5, MPI_WRF, MPI_CCLM, MPI_MM5, GFDL_WRF, and GFDL_RegCM (UKESM_WRF, UKESM_CCLM, UKESM_GRIMs, and UKESM_RegCM) for Historical, RCP2.6, and RCP8.5 simulations (Historical, SSP1-2.6, and SSP5-8.5 simulations) are ENS_CMIP5_HIS, ENS_RCP26, and ENS_RCP85 (ENS_CMIP6_HIS, ENS_SSP126, and ENS_SSP585), respectively. The average of the 12 GCM-RCM chains for the Historical simulation is named ENS_ALL_HIS. The analysis results for the present climate show only ENS_ALL_HIS, even though a future change is calculated based on the differences between ENS_RCP26 or ENS_RCP85 (ENS_SSP126 or ENS_SSP585) and ENS_CMIP5_HIS (ENS_CMIP6_HIS).

The analysis period is 25 years for the Historical simulation (1981-2005) and 30

years for RCP and SSP simulations (2071-2100). The analysis domain is the whole CORDEX-EA2 domain（Figure 2）and ocean grid points are masked as missing values. In order to analyze the characteristics of a CDNHW in each region in East Asia with its various climate zones, the whole domain is divided into several sub-domains: the Korean Peninsula (R1, 33-42°N, 125-130°E), East China (R2, 23- 42°N, 104-123°E), West China (R3, 31-42°N, 80-104°E), India (R4, 8-27°N, 74-84°E), Mongolia (R5, 42-52°N, 88-120°E), Japan (R6, 30-42°N,130-145°E), Indochina (R7, 8-23°N,92-110°E), and Northeast China (R8, 42-54°N,120-135°E). The daily maximum and minimum temperatures, which are variables used to define a CDNHW, are calculated from the highest and lowest values among the 3-hourly temperature data. In Historical, RCP, and SSP simulations, the period in which the annual maximum values are high for both the daily maximum and minimum temperatures varied between mid-April and early September in all of East Asia. That period is July and August on the Korean Peninsula and in Northeast China, but is June to August in West China, East China, and Mongolia. It is May to June in India, from July to September in Japan, and from April to June in Indochina. Therefore, the heatwave period for all of East Asia is selected as April 1 to September 30 (Figure S1).

In order to evaluate the performance of the 12 GCM-RCM chains, ERA5 reanalysis data provided by the European Centre for Medium-Range Weather Forecasts is used. Gridded data from 12 GCM-RCM chains are converted to the ERA5 grid with 0.25° resolution using a simple inverse distance weighting method to compare model results with observations.

## 2.2. Variance Scaling

In order to reduce systematic bias in each model, the daily maximum and minimum temperatures are bias-corrected using variance scaling, which is a method to correct both mean and variance in temperatures（Teutschbein and Seibert, 2012）. The observation data used for bias correction is the aforementioned ERA5 reanalysis data. In (1) to (8) below, $T_{obs}$ indicates daily maximum or minimum temperature in the observation data, while $T_{contr}$ and $T_{scen}$ indicate daily maximum or minimum temperature in Historical and RCP or SSP simulations, respectively. First, the mean of the Historical simulation is corrected by adding the difference between the long-term monthly mean observation data and the Historical simulation: $(\mu_m(T_{obs}) - \mu_m(T_{contr}))$ as seen in (1). Each mean for RCP or SSP simulations is also corrected by adding the term that is assumed to remain unvaried, even for a future climate, expressed in (2).

$$T^{*1}_{contr} = T_{contr} + \mu_m(T_{obs}) - \mu_m(T_{contr}) \quad (1)$$

$$T^{*1}_{scen} = T_{scen} + \mu_m(T_{obs}) - \mu_m(T_{contr}) \quad (2)$$

Thereafter, the mean corrected Historical simulation, $T^{*1}_{contr}$, and the RCP or SSP simulation, $T^{*1}_{scen}$, are shifted on a monthly basis to a zero mean:

$$T^{*2}_{contr} = T^{*1}_{contr} - \mu_m(T^{*1}_{contr}) \quad (3)$$

$$T^{*2}_{scen} = T^{*1}_{scen} - \mu_m(T^{*1}_{scen}) \quad (4)$$

Then, the variances of $T^{*2}_{contr}$ and $T^{*2}_{scen}$ are corrected based on the ratio of the standard deviation of $T_{obs}$ to that of $T^{*2}_{contr}$ ($\sigma_m(T_{obs})$ and $\sigma_m(T^{*2}_{contr})$):

$$T^{*3}_{contr} = T^{*2}_{contr} \cdot \left[\frac{\sigma_m(T_{obs})}{\sigma_m(T^{*2}_{contr})}\right] \quad (5)$$

$$T^{*3}_{scen} = T^{*2}_{scen} \cdot \left[\frac{\sigma_m(T_{obs})}{\sigma_m(T^{*2}_{contr})}\right] \quad (6)$$

Finally, the mean- and variance-corrected Historical simulations and the RCP or SSP simulation, $T^{*3}_{contr}$ and $T^{*3}_{scen}$, are shifted back using the value subtracted in (3)-(4):

$$T^{*}_{contr} = T^{*3}_{contr} + \mu_m(T^{*1}_{contr}) \quad (7)$$

$$T^{*}_{scen} = T^{*3}_{scen} + \mu_m(T^{*1}_{scen}) \quad (8)$$

### 2.3. Definition of concurrent daytime and nighttime heatwave

In order to analyze a CDNHW in East Asia with various climate zones, the threshold of a heatwave considering spatial and temporal climate characteristics is needed. Therefore, the daily thresholds of daytime and nighttime heatwaves are defined as the 90th percentile of the daily maximum and minimum temperatures, centered on a 31-day window (i.e., 15 days before and after a Julian day) for the reference period (1981-2005), as referred to in Russo *et al* (2015). Daytime and nighttime heatwaves are defined as a day where the daily maximum and minimum temperatures exceed the daily thresholds for daytime and nighttime heatwaves, respectively. The daily intensities of daytime and nighttime heatwaves are defined as follows (Russo *et al* 2015):

$$INT_DHW = \begin{cases} \frac{Tmax - Tmax_{25y25p}}{Tmax_{25y75p} - Tmax_{30y25p}} & (Tmax > Tmax_{25y25p}) \\ 0 & (Tmax \leq Tmax_{25y25p}) \end{cases} \quad (9)$$

$$INT_NHW = \begin{cases} \frac{Tmin - Tmin_{25y25p}}{Tmin_{25y75p} - Tmin_{25y25p}} & (Tmin > Tmin_{25y25p}) \\ 0 & (Tmin \leq Tmin_{25y25p}) \end{cases} \quad (10)$$

In equation (9), $Tmax$ is the daily maximum temperature on day when a daytime heatwave occurs, while $Tmax_{25y25p}$ and $Tmax_{25y75p}$ are the 25th and 75th percentiles, respectively, for annual maximum values from daily maximum temperatures during the reference period. In equation (10), $Tmin$ is the daily minimum temperature on day when a nighttime heatwave occurs, while $Tmin_{25y25p}$ and $Tmin_{25y75p}$ are the 25th and 75th percentiles, respectively, for annual maximum values from daily minimum temperatures during the reference period. When INT_DHW is greater than 0 and INT_NHW is less than 0, it is defined as an independent daytime heatwave (IDHW), but when INT_DHW is less than 0 and INT_NHW is greater than 0, it is an independent nighttime heatwave (INHW). When both INT_DHW and INT_NHW are greater than 0, it is a CDNHW. The daily intensity of a CDNHW (INT_CDNHW) is defined as the sum of INT_DHW and INT_NHW on the day the CDNHW occurs. A day with the daily maximum or minimum temperature is higher than daily threshold of daytime or nighttime heatwave, although daily intensity is zero, is considered as a hot day or hot night, not as a heatwave day.

## 3. Results

First, the performance of ENS_ALL_HIS in simulating the daily maximum and minimum temperatures used to define a CDNHW is evaluated over the eight sub-domains. Figure 3 presents Taylor diagrams that show the temporal standard deviations for ENS_ALL_HIS normalized with ERA5 (NSD), and the spatial correlation coefficients (SCCs) between ENS_ALL_HIS and ERA5. The closer to the point marked REF (1.0), the more similar the performance of ENS_ALL_HIS to ERA5. The SCC and NSD for daily maximum temperatures derived from ENS_ALL_HIS before bias correction (Tmax_ORG) appear within the ranges 0.82 to 0.98 and 0.82 to 1.05, respectively, and those for the daily minimum temperatures derived from ENS_ALL_HIS before bias correction (Tmin_ORG) appear within the ranges 0.89 to 0.99 and 0.90 to 1.00, respectively. This means simulated daily maximum and minimum temperatures in ENS_ALL_HIS are similar to those of ERA5 in East Asia, although there are some variations depending on the region. The SCC for corrected daily maximum and minimum temperatures derived from ENS_ALL_HIS via the variance scaling method (Tmax_VS and Tmin_VS) appear within the ranges 0.97 to 1.00 and 0.98 to 1.00, respectively. Both NSDs for Tmax_VS and Tmin_VS are close to 1.0. Likewise, in twelve GCM-RCM chains, simulated daily maximum and minimum temperatures are also similar to those of ERA5 in East Asia and Tmax_VS and Tmin_VS are closer to the REF than Tmax_ORG and Tmin_ORG (Figure S2). In addition, through the probability distribution functions (Figure S3), we can see that the distribution ranges and shapes for Tmax_VS and Tmin_VS are more similar to those of ERA5 than Tmax_ORG and Tmin_ORG. Hence, CDNHWs were analyzed using Tmax_VS and Tmin_VS in this study.

We investigate the temporal and spatial characteristics of daily maximum (Tmax) and minimum temperatures (Tmin) (Figures 4 and 5). In ENS_ALL_HIS, the Tmax and Tmin are the highest from April to June for India and Indochina; and in July or August for the Korean Peninsula, China, Mongolian, and Japan. In India and Indochina, the Tmax and Tmin are higher than other regions, while the monthly variability of those are relatively small (Figure S4). Figure S5 show the spatial distributions of mean Tmax and Tmin during the heatwave period in East Asia from April to September derived from ERA5, ENS_CMIP5_HIS, and ENS_CMIP6_HIS, as well as the bias compared with ERA5. The magnitude of bias for mean Tmax and Tmin derived from ENS_CMIP5_HIS is smaller than that derived from ENS_CMIP6_HIS, which may be because ENS_CMIP6_HIS is forced by only one GCM, while ENS_CMIP5_HIS is forced by three GCMs. However, the difference between ENS_CMIP5_HIS and ENS_CMIP6_HIS is negligible over East Asia, except for the mean Tmax in India. These results indicate that ENS_SSP126 and ENS_SSP585 can help to confirm that future projections of heatwaves under SSP126 and SSP585 scenarios support those under RCP2.6 and RCP8.5 scenarios. The temporal and spatial characteristics of a daytime and nighttime heatwave threshold is similar to the daily maximum and daily minimum temperatures (Figure S6).

In ENS_RCP26, ENS_RCP85, ENS_SSP126 and ENS_SSP585, the Tmax and Tmin increase in all regions of East Asia, especially, more increase in high-latitude region and more increase during the period when temperatures are already high in ENS_ALL_HIS. In addition, the Tmin increase more than Tmax in East Asia expect for East China and Indochina. The average of Tmax/Tmin during the period from April to September increase from 22.5°C/13.9°C (ENS_ALL_HIS) to 23.6°C/15.1°C (ENS_RCP26), 26.3°C/17.7°C (ENS_RCP85), 25.6°C/17.0°C (ENS_SSP126) and 29.5°C/21.0°C (ENS_SSP585) in East Asia. Therefore, the increases are large in the following order: ENS_SSP585, ENS_RCP85, ENS_SSP126, and ENS_RCP26. SSP5-8.5 and RCP8.5 (SSP1-2.6 and RCP2.6) scenarios represent the high-emission (low-emissions) scenarios with the same radiative forcing of 8.5 W/m2 (2.6 W/m2) by 2100, but the CO2 emissions at the end of the 21st century in SSP5-8.5 (SSP1-2.6) scenario is approximately 21.0% (6.0%) higher than that in RCP8.5 (RCP2.6) scenario. Therefore, the Tmax and Tmin under SSP scenarios are higher than those under corresponding RCP scenarios.

Future changes in IDHWs, INHWs, and CDNHWs due to the increase in daily maximum and minimum temperatures were analyzed. Figure 6 presents pie charts showing the occurrence rates for IDHWs, INHWs, and CDNHWs during the heatwave period in East Asia (OR_IDHW, OR_INHW, and OR_CDNHW). In ERA5, OR_IDHW, OR_INHW, and OR_CDNHW appear within the ranges 1.2 to 1.5%, 0.9 to 1.7%, and 0.5 to 1.2% in the eight sub-regions, which means that CDNHWs occur rarely than INHW and IDHW. Although ENS_ALL_HIS tends to overestimate or underestimate OR_IDHW, OR_INHW, and OR_CDNHW, depending on the sub-region, the absolute values in the bias of those area averages in East Asia are very small (0.1%, 0.5%, and 0.3%, respectively). This means OR_IDHW, OR_INHW, and OR_CDNHW derived from ENS_ALL_HIS and ERA5 are very similar. In ENS_RCP26, ENS_RCP85, ENS_SSP126, and ENS_SSP585, OR_IDHW, OR_INHW, and OR_CDNHW increase in all regions of East Asia. In addition, the regional differences in OR_IDHW, OR_INHW, and OR_CDNHW are large compared to ENS_ALL_HIS because the degree of the increase vary depending on the region. In the current climate, the occurrence rate of a CDNHW area-averaged in East Asia is 0.9%, which is lower than for IDHWs (1.4%) and INHWs (1.6%). In the future, the occurrence rates of CDNHWs area-averaged in East Asia will increase to 3.7% (RCP2.6), to 20.5% (RCP8.5), to 15.6% (SSP1-2.6), and to 45.7% (SSP5-8.5), and will be higher in Indochina than in other regions. Even, CDNHW more occurs than IDHW and INHW in Mongolia and Japan (RCP2.6), on the Korean Peninsula, and in East and West China, India, Mongolia, and Japan (RCP8.5), on the Korean Peninsula, and in West China, Mongolia, Japan, and Northeast China (SSP1-2.6), and in all regions of East Aisa (SSP5-8.5).

In order to examine the characteristics of a CDNHW, first, the period in which the CDNHW mainly occurs in each sub-domain is analyzed. Box plots presented in Figure 7 show the climatology of the Julian day when CDNHW occurs first and last (first date and last date of CDNHW) and the climatology of CDNHW occurrence period, which is calculated as the difference between last date and first date of CDNHW, for grids in the eight sub-domain. In ENS_ALL_HIS, CDNHWs mainly occurs from late July to mid-August on the Korean Peninsula, from mid-July to early August in East and West China, from early-May to early June in India, from early to late July in Mongolia, from late July to mid-August in Japan, from late April to early June in Indochina, and from mid to late July in Northeast China. The area-averaged occurrence period of CDNHW in East Asia is about 10.9 days, which is relatively long in East China, Mongolia, Japan, and Indochina. In ENS_RCP26, ENS_RCP85, ENS_SSP126 and ENS_SSP585, the first date of CDNHW is earlier and last date of CDNHW is later than in ENS_ALL_HIS, respectively (Table S2 and Table S3). Therefore, the occurrence period of CDNHW area-averaged in East Asia extends to about 23.2 days (RCP2.6), 66.0 days (RCP8.5), 59.3 days (SSP1-2.6), and 110.2 days (SSP5-8.5). In other words, CDNHWs occur even when CDNHWs do not usually occur in ENS_ALL_HIS. The occurrence period is relatively longer in East and West China, Mongolia, and Indochina (RCP2.6), in East and West China and Indochina (RCP8.5 and SSP5-8.5) and in East and West China, Mongolia, and Indochina (SSP1-2.6), compared to other regions.

Figure 8 presents the frequency of CDNHWs of various durations. In ENS_ALL_HIS (Figure 8a), the duration of a CDNHW is mostly from one to three days in East Asia. Although the frequency of all CDNHWs is high in Mongolia and Japan, CDNHWs lasting more than four days occur more in India and Japan than other regions. In ENS_RCP26, ENS_RCP85, ENS_SSP126 and ENS_SSP585, the frequency of all CDNHWs increases, with a CDNHW lasting fewer than three days being the most common, as seen in ENS_ALL_HIS. In particular, long-lasting CDNHWs, which rarely occurs in ENS_ALL_HIS, occur. In ENS_RCP26 (Figure 8b), CDNHWs lasting more than seven days occur on the Korean Peninsula, and in East China, India, Japan, and Indochina. In ENS_RCP85 and ENS_SSP126 (Figures 8c,d), CDNHWs lasting more than seven days occur in all regions of East Asia, and those lasting more than ten days occur in most regions of East Asia except for Mongolia and Northeast China. In ENS_SSP585 (Figure 8e), CDNHWs lasting more than ten days occurrs in all regions of East Asia. Therefore, the proportion of CDNHWs lasting more than ten days, compared to the frequency of all CDNHWs, in East Asia will increase to 1.2% (RCP2.6), 7.2% (RCP8.5), 6.1% (SSP1-2.6), and 17.3% (SSP5-8.5) from 0.2% (Historical). These long-lasting CDNHWs will occur more frequently in India and Indochina under RCP2.6 and RCP8.5 scenarios, in Japan and Indochina under SSP1-2.6 scenario, and in India and Japan under SSP5-8.5 scenario. Among the high-emission scenarios, CDNHW lasting more than ten days is three times more frequent in SSP5-8.5 scenario than RCP8.5 scenario.

Finally, we investigated future changes in the intensity of CDNHWs due to increases in the occurrence periods and the number of days they last. Figures 9 and S7 show the spatial distributions for accumulated intensity of CDNHWs derived from ENS_ALL_HIS, ENS_RCP26, ENS_RCP85, ENS_SSP126, and ENS_SSP585 and their differences. The spatial extent of CDNHW, which is calculated as a percentage of CDNHW area to the total sub-domain area, is shown in Figure S8. In ENS_ALL_HIS, the accumulated intensity of CDNHW from April to September is relatively high in West China (except for the Tibetan Plateau), in some regions of East China, India, and in Thailand and Cambodia in Indochina. The period when intensity of CDNHW is the highest over a broad area is May in India and Indochina, and July or August on the Korean Peninsula, in China, in Mongolia, and in Japan.

In ENS_RCP26, ENS_RCP85, ENS_SSP126 and ENS_SSP585, not only the intensity of CDNHW more increases but also the spatial and temporal extent of the CDNHW is more expanded. The intensity of CDNHW increases in all regions of East Asia and that increases more in East China, West China, India and in Indochina, where the already more affected by CDNHW in ENS_ALL_HIS. While, the intensity of CDNHW derived from ENS_RCP26 decrease in the eastern part of East China, the western part of West China, and Northeast China. The increase in the intensities of CDNHWs is larger in the order of SSP5-8.5, RCP8.5, SSP1-2.6, and RCP2.6 scenarios. In particular, the increase is four times stronger under SSP5-8.5 scenario than RCP8.5 scenario. In addition, CDNHW occurs even in the period when the CDNHW rarely occurs in ENS_ALL_HIS (in May, June, and September on the Korean Peninsula, in April in East China, in April and May in West China, from July to September in India, in April and September in Mongolia, from April to June in Japan, in September in Indochina, and in April, May, and September in Northeast China).

## 4. Discussion and conclusion

The adverse effects of extreme heat are expected to increase due to global warming. In particular, CDNHWs will have negative impacts not only on ecosystems and human health but across industries in all countries. Although quantitatively analyzing future changes in CDNHWs (as well as daytime or nighttime heatwaves) is important, few studies have analyzed future changes in CDNHWs over East Asia. In this study, CDNHWs over the whole of East Asia by the end of the 21$^{st}$ century (2071-2100) are projected under RCP2.6, RCP8.5, SSP1-2.6, and SSP5-8.5 scenarios using 3-hourly temperature data with a 25km horizontal resolution produced by 12 GCM-RCM chains participating in the CORDEX-EA2 project.

To evaluate the performance of GCM-RCM chains in simulating daily maximum and minimum temperatures, NSD, SCC, and probability distribution functions were examined. As a result, the temporal and spatial climate characteristics of bias-corrected daily maximum and minimum temperatures under the variance scaling method were more similar to those of ERA5 than daily maximum and minimum temperatures before bias correction. In Historical simulation, the occurrence rate for CDNHWs from April to September (area-averaged in East Asia) is 0.9%, which is very low in all regions of East Asia. A CDNHW mainly occurs from July to August on the Korean Peninsula, in China, in Mongolia, and in Japan, and occurs from April to June in India and Indochina. The occurrence period for CDHNWs from first to last onset is about one to two weeks, and the duration of a CDNHW is usually less than three days. The period when the intensity of a CDNHW is the highest and when it covers the widest area is in July or August on the Korean Peninsula, in China, in Mongolia, and in Japan, and is in May for India and Indochina. The accumulated intensity of a CDNHW from April to September is higher in West China, except for the Tibetan Plateau, and in East China, India, as well as Thailand and Cambodia in Indochina, compared to other regions.

By the end of the 21st century, the occurrence period for CDHNWs will be extended to three weeks (RCP2.6), to two month (RCP8.5 and SSP1-2.6), and to three months (SSP5-8.5), and the occurrence rates for CDNHWs will increase to 3.7% (RCP2.6), 20.5% (RCP8.5), 15.6% (SSP1-2.6) and 45.7% (SSP5-8.5), as the daily maximum and minimum temperatures from April to September will increase in all regions of East Asia. Therefore, it is expected that the intensity of CDNHWs will increase, and the spatial extent of CDNHW will be expanded. The period when the intensity of a CDNHW is the highest and when it covers the widest area will be expected to be the same period as in Historical simulation. CDNHWs lasting less than three days are most common, but the frequency of CDNHWs lasting more than ten days will increase significantly. The occurrence period and duration of CDNHWs will be longest and occurrence rate and intensity of CDNHWs will be highest in Indochina, and those are longer or higher in East China, West China, and India, compared to other regions. More frequent, intense, and prolonged CDNHWs occur over a broad area in high-emission scenarios, particularly in SSP5-8.5 scenario, because future warming in East Asia is more pronounced under SSP scenarios than under corresponding RCP scenarios.

The physical or dynamical mechanisms responsible for the increase in intensity and frequency of CDNHWs in each region of East Asia should be explored in-depth in further study. Recently, many study attach the attention to CDNHWs, but underlying mechanism of CDNHWs remains unclear. Very few studies show that the persistent anticyclonic circulation is mainly associated with concurrent daytime and nighttime heatwaves in East Asia regions (Li et al 2021; Luo et al 2022 for China, Wang et al 2020 in Northern hemisphere). The anomalous anticyclone contributes to surface warming through adiabatic heating, additional solar radiation absorption, and offsetting the nighttime radiative cooling (Lee and Lee 2016; Hong et al 2018; Choi and Lee 2019; Im et al 2019; Kim et al 2019; Yoon et al 2020; Noh et al 2021; Seo et al 2021; Yoon et al 2021 for Korean Peninsula, Park et al 2012; Freychet et al 2017; Luo and Lau 2017; Wang et al 2019; Hong et al 2020 for China, Ratnam *et al* 2016; Rohini et al 2016; Sandeep and Prasad 2018; Joshi et al 2020 for India, Erdenebat and Sato 2016 for Mongolia, Noh et al 2021; Nishi et al 2022 for Japan Luo and Lao 2018 for Indochina Peninsula). We carried out the composite analysis of the daytime and nighttime geopotential height at 500hPa (Z500) anomalies for CDNHW events using 6-hourly Z500 data. Considering local time, Z500 at 06 UTC and 18 UTC were used to composite mean for daytime and nighttime hour, respectively. Figures S9 and S10 show that the positive Z500 anomalies persist day and night over the regions with CDNHW events in Historical simulation. In RCP and SSP scenarios, the magnitude of daytime and nighttime Z500 anomaly against reference period climatology increases, which may be one of the reasons for increase in CDNHWs.

Recently, the CDNHWs are receiving scientific attention, but the underlying mechanism related to that has been poorly understood. Very a few studies show that the persistent anticyclonic circulation is mainly associated with concurrent daytime and nighttime heatwaves in East Asia regions (Li et al 2021; Luo et al 2022 for China, Wang et al 2021 in Northern hemisphere). In order to understand the underlying mechanism associated with CDNHW in East Asia, first of all, we carried out the composite analysis of the daytime (06UTC) and nighttime (18UTC) geopotential height at 500hPa (Z500) anomalies for CDNHW events. Figure 10 shows the composites of daytime or nighttime Z500 anomalies for each of IDHW, INHW, and CDNHW in Historical simulation. It is founded that positive Z500 anomalies is dominant during the day and night when IDHW and INHW occurs, and that persist day and night when CDNHW occurs over each sub-region. In general, under the anomalous anticyclone conditions, adiabatic heating due to sinking motion and additional solar radiation absorption due to reduction of cloud cover contribute to surface warming during the daytime. In addition, warm advection caused by anomalous wind, and, the reduction of nighttime radiative cooling due to anomalous moisture advection, particularly in coastal areas as well as the adiabatic heating elevates surface temperature during the night (Lee and Lee 2016; Hong *et al* 2018; Choi and Lee 2019; Im *et al* 2019; Kim *et al* 2019; Yoon *et al* 2020; Noh *et al* 2021; Seo *et al* 2021; Yoon *et al* 2021 for Korean Peninsula, Park *et al* 2012; Freychet *et al* 2017; Luo and Lau 2017; Wang *et al* 2019; Hong *et al* 2020 for China, Ratnam *et al* 2016; Rohini *et al* 2016; Sandeep and Prasad 2018; Joshi *et al* 2020 for India, Erdenebat and Sato 2016 for Mongolia, Noh *et al* 2021; Nishi *et al* 2022 for Japan Luo and Lao 2018 for Indochina Peninsula). In RCP and SSP scenarios, the magnitude of daytime and nighttime Z500 anomaly against reference period climatology is much higher than the magnitude of that in Historical simulation, which may be one of the reasons for increase in CDNHWs (Figure S9 and Figure S10). The physical or dynamical mechanisms contributing to the increase in intensity and frequency of CDNHWs in each region of East Asia should be explored in-depth in further study.

In summary, the frequency and intensity of CDNHWs, which occur rarely during the heatwave periods in the present climate, will increase significantly by the end of the 21st century over the whole of East Asia. These increases are expected to be greater in Indochina, India, East China, and West China, compared to other regions.

**Acknowledgements**

This work was carried out with the support of the Korea Meteorological Administration Research and Development Program under Grant KMI2020-01411.

## Figure legend section

**Figure 1.** Twelve GCM-RCM chains.

**Figure 2.** The CORDEX-East Asia phase 2 domain and topography (unit: m). Boxed areas indicate the eight sub-domains (R1-R8).

**Figure 3.** Taylor diagrams of the 25-year (1981-2005) mean (a) daily maximum temperature (Tmax) and (b) daily minimum temperature (Tmin) during the heatwave period in East Asia derived from ENS_ALL_HIS. Radial axes show the temporal standard deviation of ENS_ALL_HIS normalized with ERA5, and the arcs denote the spatial correlation coefficients between ENS_ALL_HIS and ERA5. The REF point (1.0) denotes where ENS_ALL_HIS exactly agrees with ERA5. In the legend, _ORG and _VS, respectively, indicate non-bias corrected and bias-corrected Tmax or Tmin.

**Figure 4.** Spatial distributions of monthly mean daily maximum temperatures during the heatwave period from April to September in East Asia derived from (top) ENS_ALL_HIS, and the future changes derived from (second from the top to the bottom) ENS_RCP26, ENS_RCP85, ENS_SSP126, and ENS_SSP585 (Unit: $^{o}$C). The grid points where the difference is significant at the 95% confidence level based on the Student`s t-test are marked with green dots.

**Figure 5.** Same as Figure 4 but shows daily minimum temperatures.

**Figure 6.** Pie charts showing independent daytime heatwave (IDHW), independent nighttime heatwave (INHW), and concurrent daytime and nighttime heatwave (CDNHW) occurrence rates during the heatwave periods over all of East Asia and the eight sub-regions, derived from (top row) ERA5, and (subsequently from top down) ENS_ALL_HIS, ENS_RCP26, ENS_RCP85, ENS_SSP126, and ENS_SSP585 (Unit: %).

**Figure 7.** Box plots showing the climatology of the (a) first date and (b) last date when CDNHWs occur (FD_CDNHW and LD_CDNHW) and (c) the climatology of the CDNHW occurrence period (OP_CDNHW) for grids in the eight sub-domains. Black, green, orange, blue, and red lines denote the SD_CDNHW, ED_CDNHW and OP_CDNHW derived from ENS_ALL_HIS, ENS_RCP26, ENS_RCP85, ENS_SSP126, and ENS_SSP585, respectively. Each box denotes the range from the 25th percentile to the 75th percentile. The line in the middle of the box denotes the median. The upper and lower whiskers indicate maximum and minimum values.

**Figure 8.** Frequency of concurrent daytime and nighttime heatwaves (CDNHWs) of various durations over the eight sub-regions for (a) ENS_ALL_HIS, (b) ENS_RCP26, (c) ENS_RCP85, (d) ENS_SSP126, and (e) ENS_SSP585.

**Figure 9.** Spatial distributions for monthly accumulated intensity of concurrent daytime and nighttime heatwaves (CDNHWs) during the heatwave period in East Asia from April to September for (top to bottom) ENS_ALL_HIS, ENS_RCP26, ENS_RCP85, ENS_SSP126, and ENS_SSP585.

**Figure 10**. Composites of geopotential height at 500hPa anomalies during daytime for IDHWs (top row) and CDNHWs (second row), and during nighttime for INHWs (third row) and CDNHWs (bottom row) over eight sub-regions (Unit: m) in Historical simulation. The grid points where the anomaly is significant at the 95% confidence level based on the Student`s t-test are marked with black dots.

# Concurrent daytime and nighttime heatwaves in the late 21st century over the CORDEX-East Asia phase 2 domain using Multi-GCM and Multi-RCM Chains

**Young-Hyun Kim[1], Joong-Bae Ahn[1*], Myoung-Seok Suh[2,] Dong-Hyun Cha[3], Eun-Chul Chang[2], Seung-Ki Min[3], Younhg-Hwa Byun[5] and Jin-Uk Kim[5]**

[1]Department of Atmospheric Sciences, Pusan National University, Busan, Korea

[2]Department of Atmospheric Science, Kongju National University, Gongju, South Korea

[3]Department of Urban and Environmental Engineering, Ulsan National Institute of Science and Technology, Ulsan, South Korea

[4]Division of Environmental Science and Engineering, Pohang University of Science and Technology, Pohang, South Korea

[5]Climate Change Research Team, National Institute of Meteorological Sciences

*Corresponding Author : Joong-Bae Ahn (E-mail : jbahn@pusan.ac.kr)

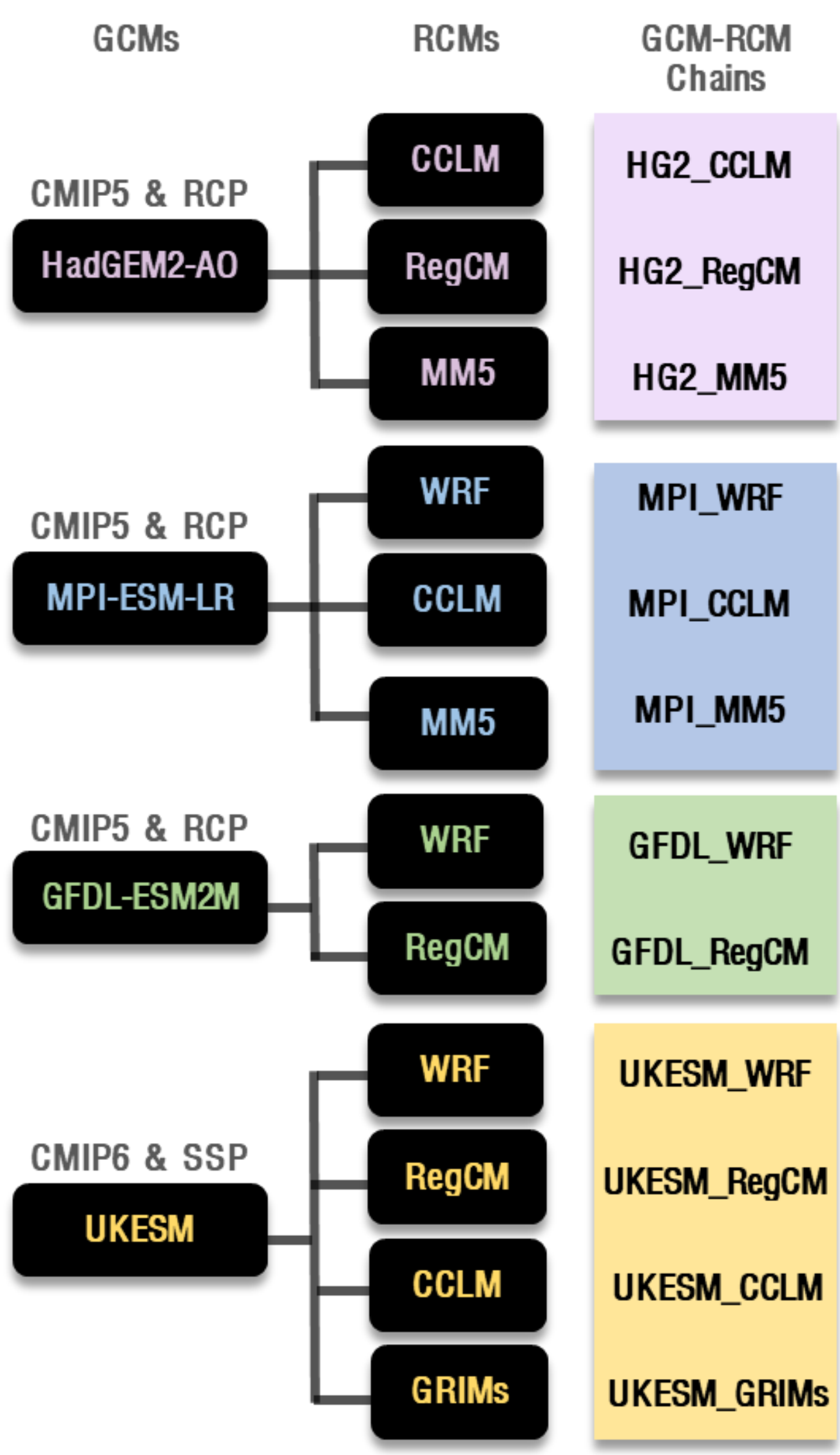

**Figure 1.** Twelve GCM-RCM chains.

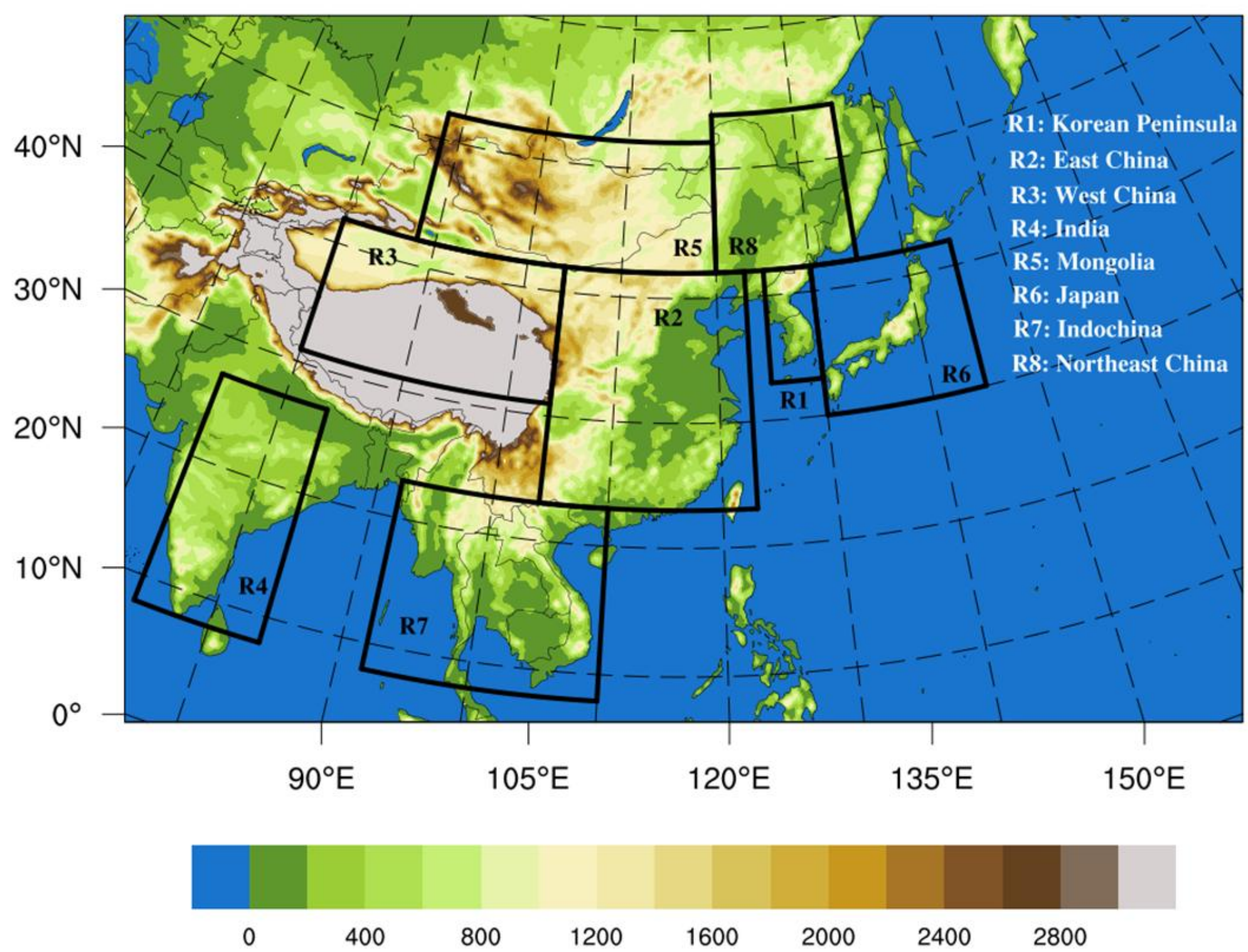

**Figure 2.** The CORDEX-East Asia phase 2 domain and topography (unit: m). Boxed areas indicate the eight sub-domains (R1-R8).

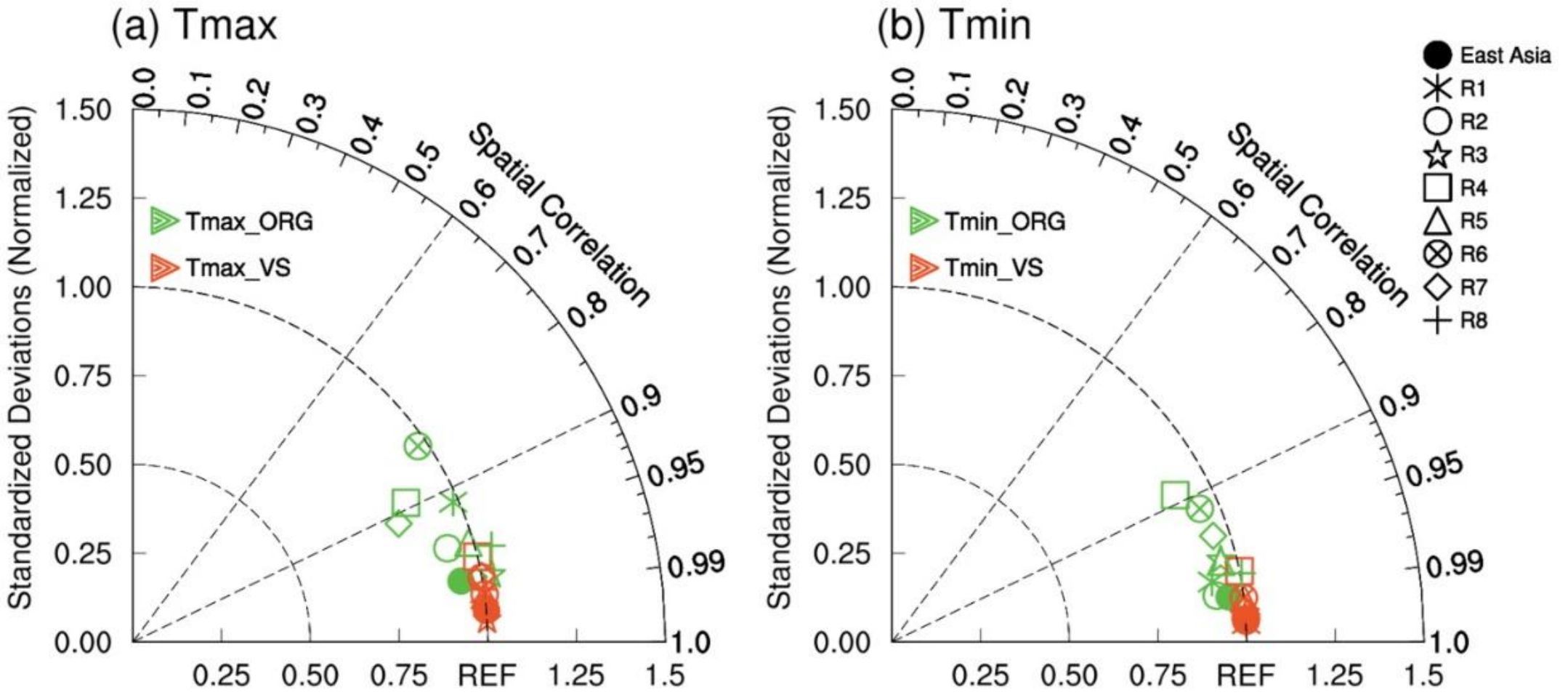

**Figure 3.** Taylor diagrams of the 25-year (1981-2005) mean (a) daily maximum temperature (Tmax) and (b) daily minimum temperature (Tmin) during the heatwave period in East Asia derived from ENS_ALL_HIS. Radial axes show the temporal standard deviation of ENS_ALL_HIS normalized with ERA5, and the arcs denote the spatial correlation coefficients between ENS_ALL_HIS and ERA5. The REF point (1.0) denotes where ENS_ALL_HIS exactly agrees with ERA5. In the legend, _ORG and _VS, respectively, indicate non-bias corrected and bias-corrected Tmax or Tmin.

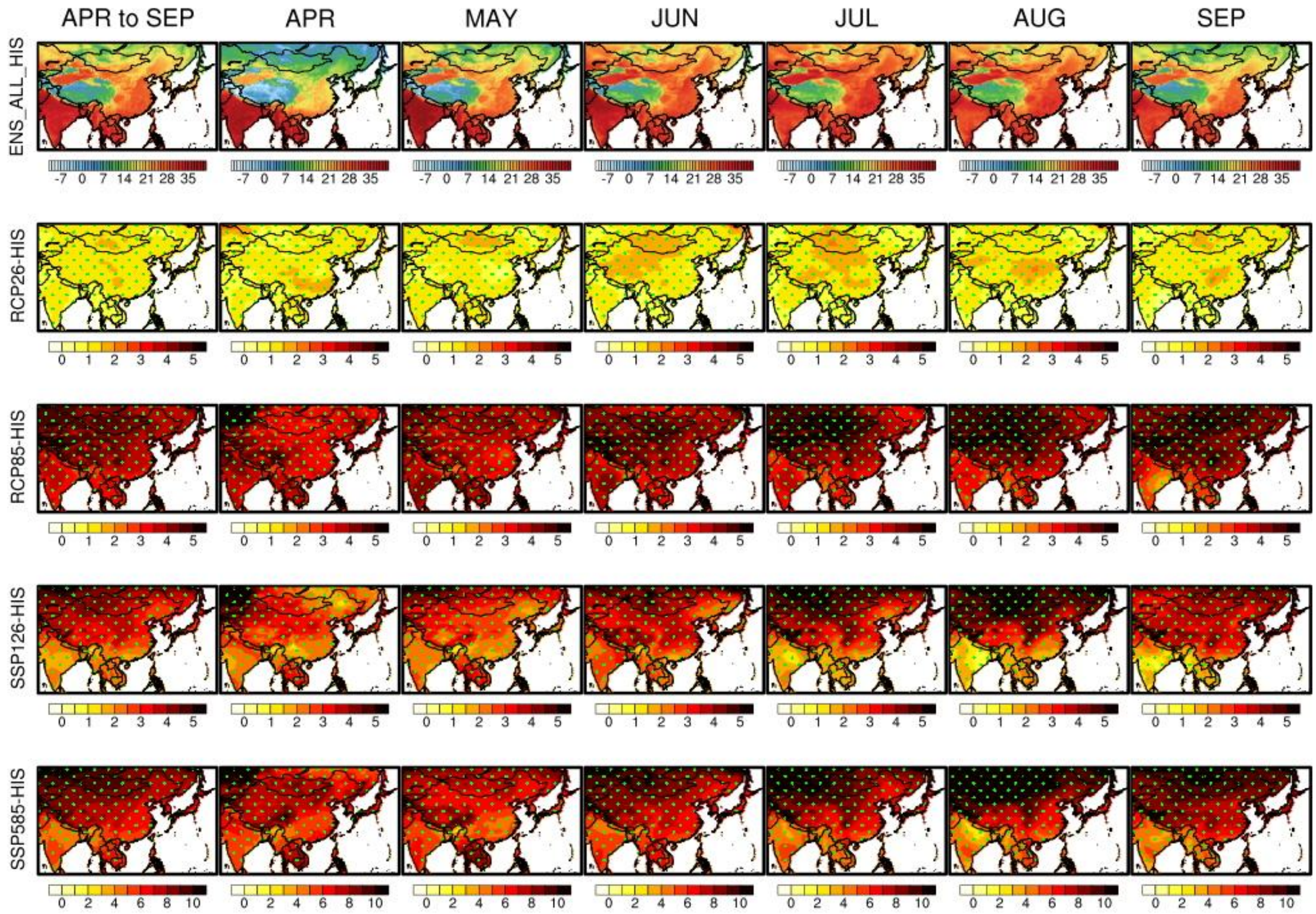

**Figure 4.** Spatial distributions of monthly mean daily maximum temperatures during the heatwave period from April to September in East Asia derived from (top) ENS_ALL_HIS, and the future changes derived from (second from the top to the bottom) ENS_RCP26, ENS_RCP85, ENS_SSP126, and ENS_SSP585 (Unit: ºC). The grid points where the difference is significant at the 95% confidence level based on the Student`s t-test are marked with green dots.

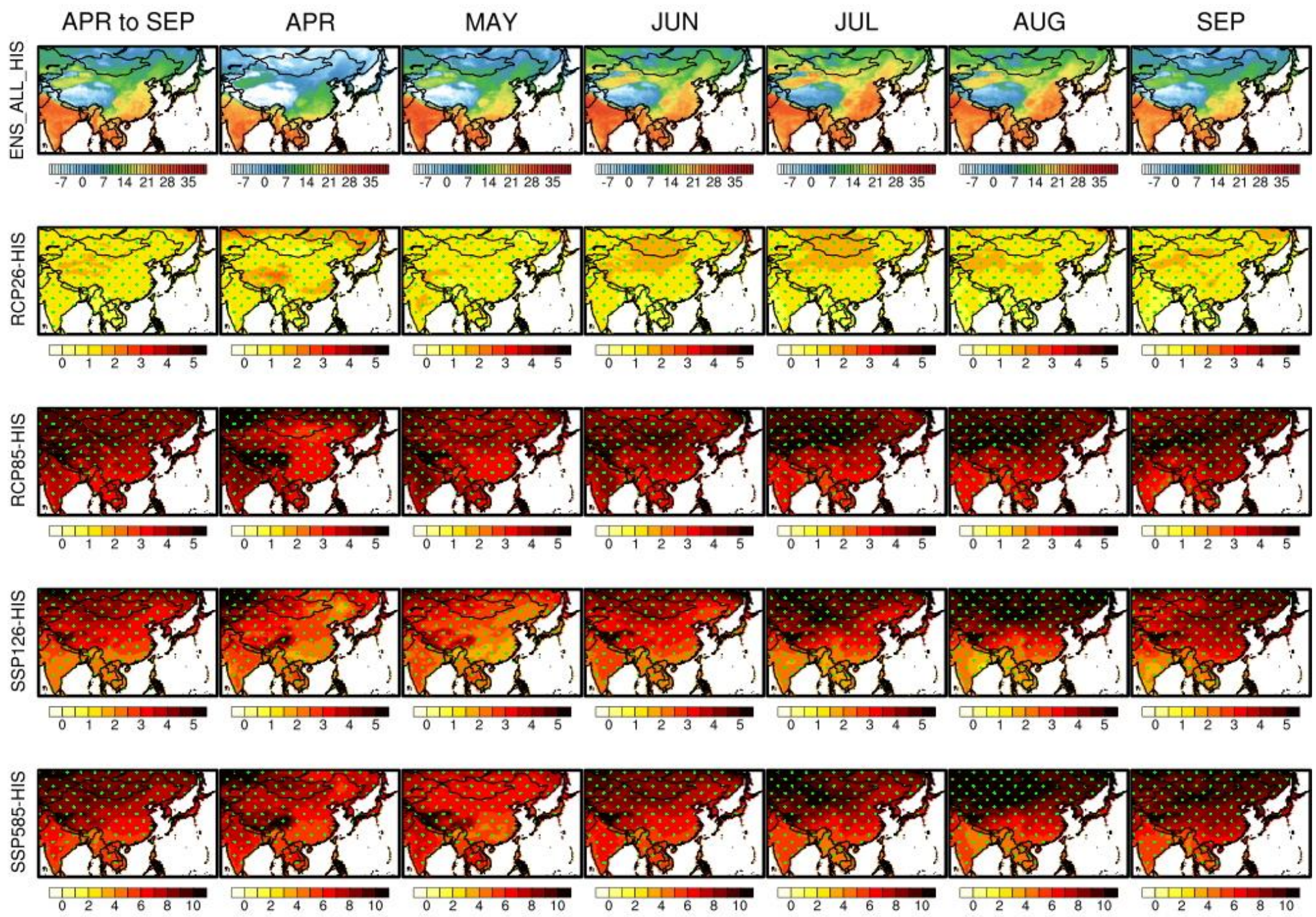

**Figure 5.** Same as Figure 4 but shows daily minimum temperatures.

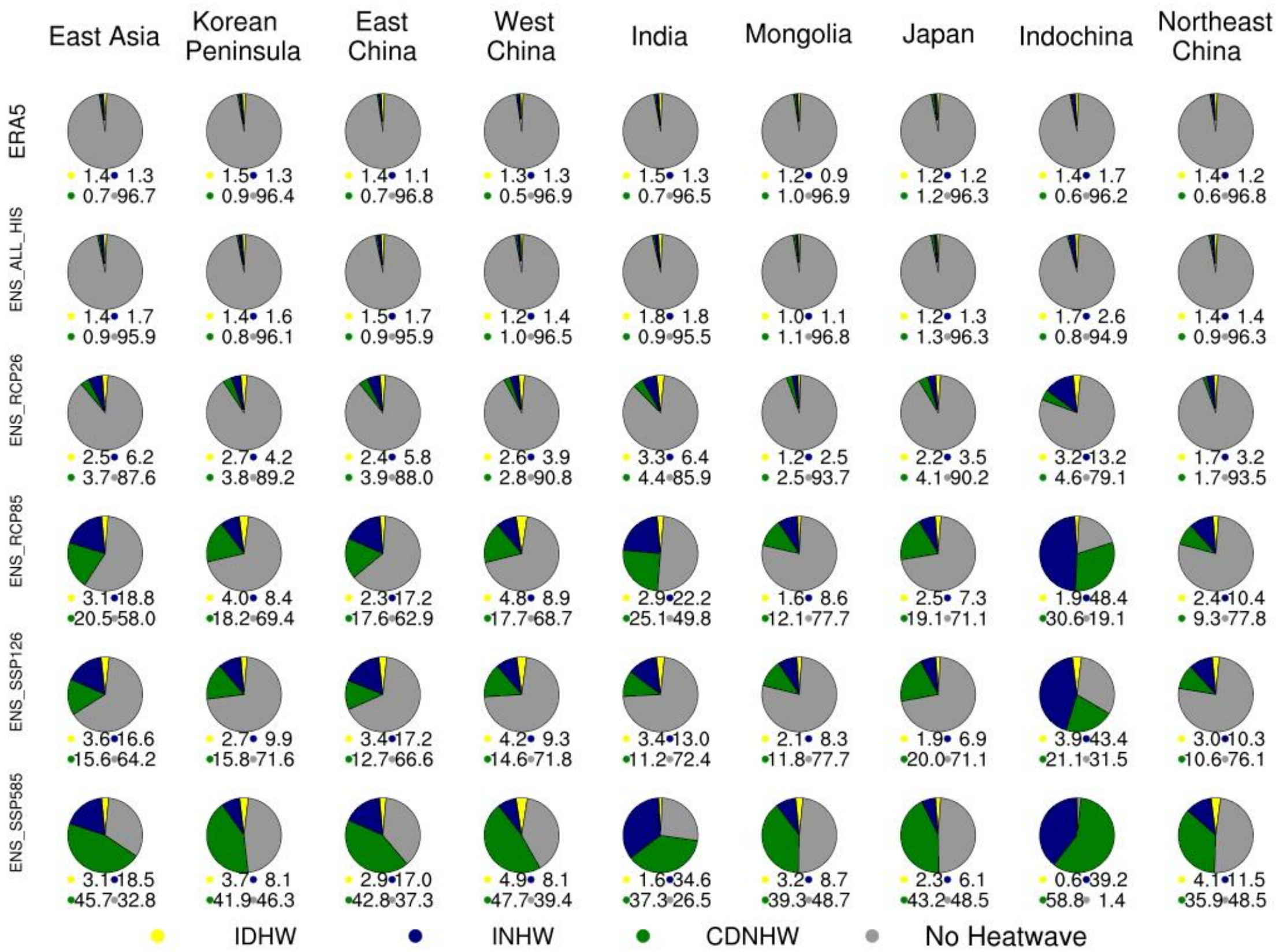

**Figure 6.** Pie charts showing independent daytime heatwave (IDHW), independent nighttime heatwave (INHW), and concurrent daytime and nighttime heatwave (CDNHW) occurrence rates during the heatwave periods over all of East Asia and the eight sub-regions, derived from (top row) ERA5, and (subsequently from top down) ENS_ALL_HIS, ENS_RCP26, ENS_RCP85, ENS_SSP126, and ENS_SSP585 (Unit: %).

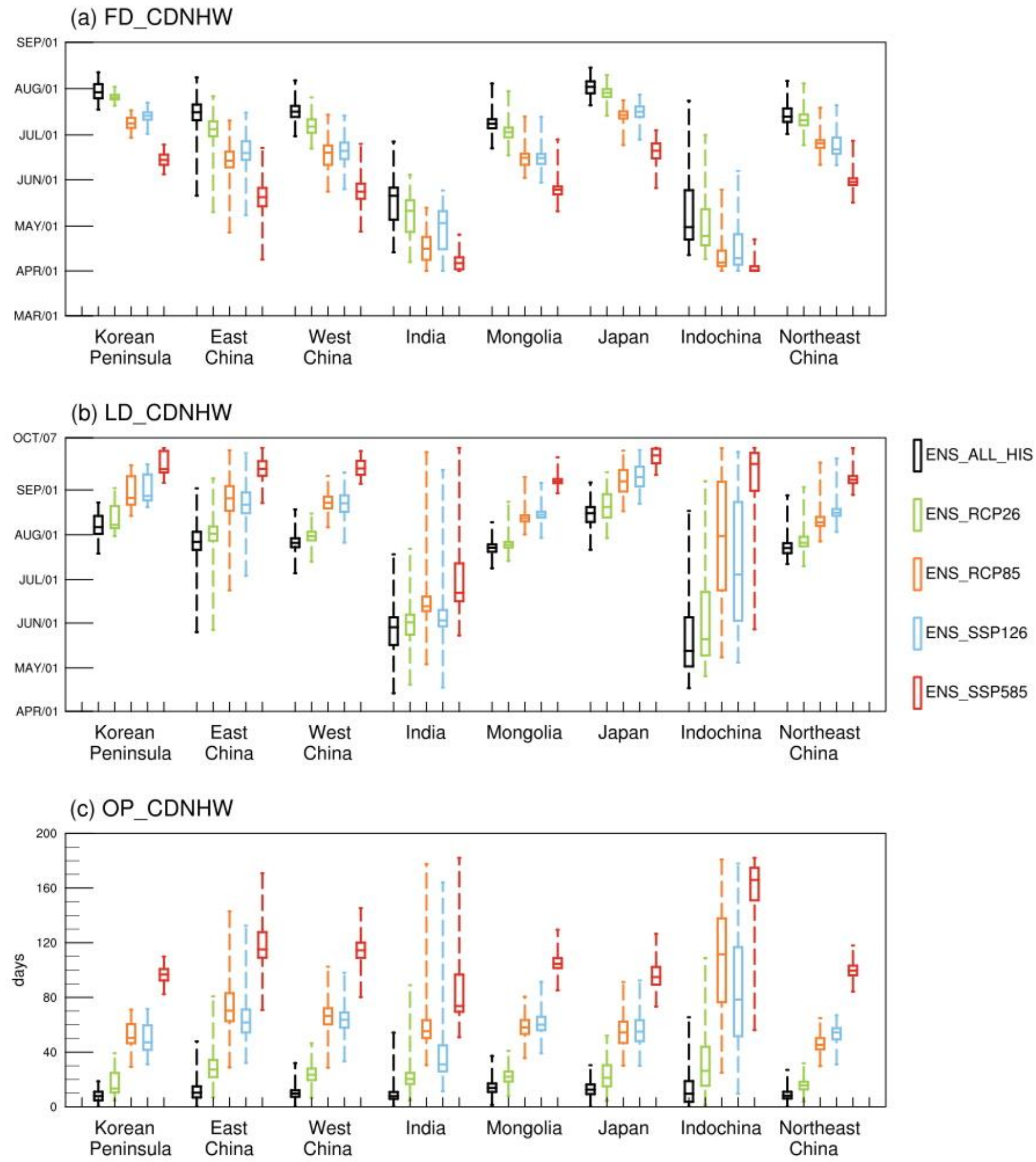

**Figure 7.** Box plots showing the climatology of the (a) first date and (b) last date when CDNHWs occur (FD_CDNHW and LD_CDNHW) and (c) the climatology of the CDNHW occurrence period (OP_CDNHW) for grids in the eight sub-domains. Black, green, orange, blue, and red lines denote the SD_CDNHW, ED_CDNHW and OP_CDNHW derived from ENS_ALL_HIS, ENS_RCP26, ENS_RCP85, ENS_SSP126, and ENS_SSP585, respectively. Each box denotes the range from the 25th percentile to the 75th percentile. The line in the middle of the box denotes the median. The upper and lower whiskers indicate maximum and minimum values.

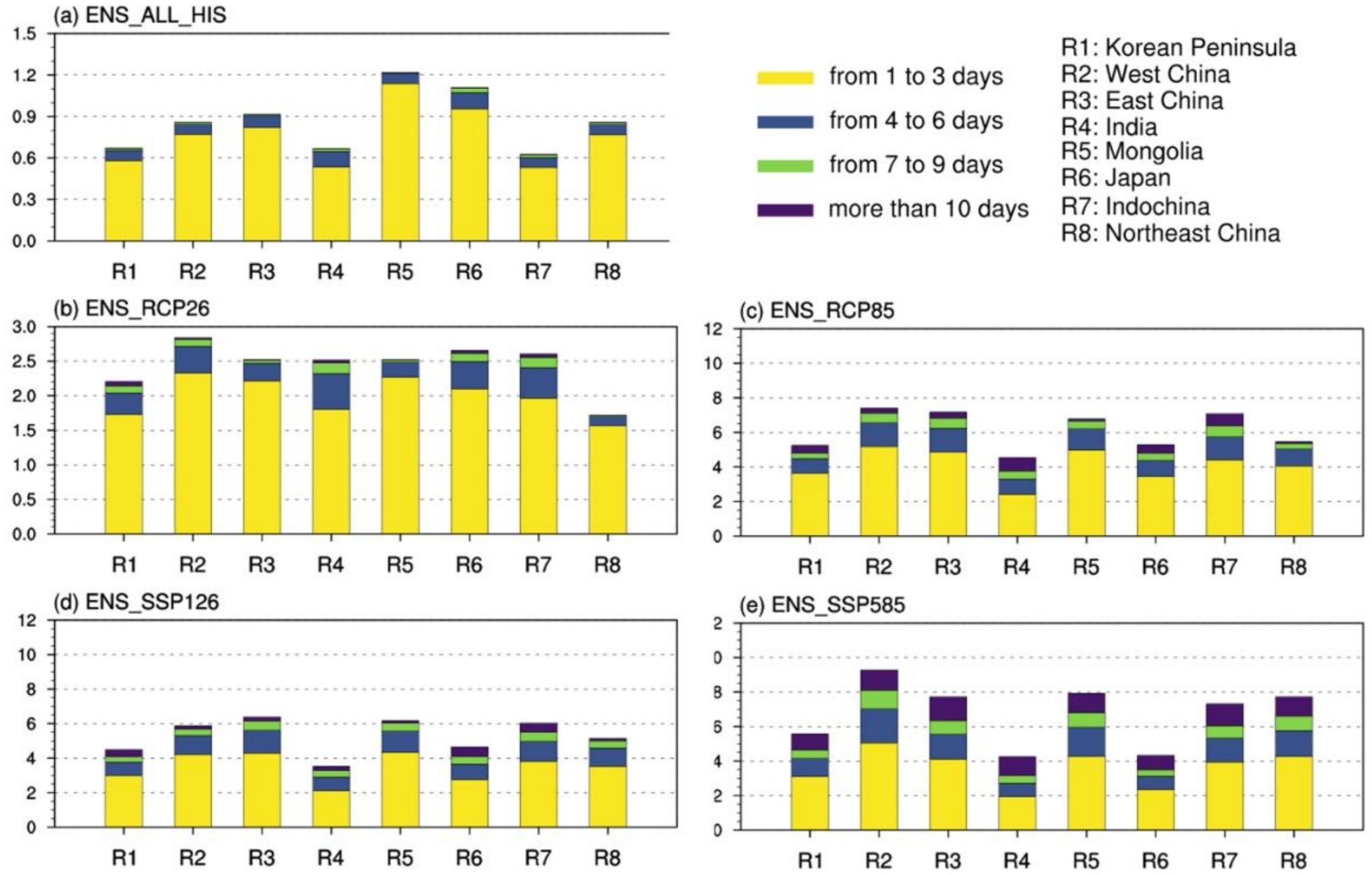

**Figure 8.** Frequency of concurrent daytime and nighttime heatwaves (CDNHWs) of various durations over the eight sub-regions for (a) ENS_ALL_HIS, (b) ENS_RCP26, (c) ENS_RCP85, (d) ENS_SSP126, and (e) ENS_SSP585.

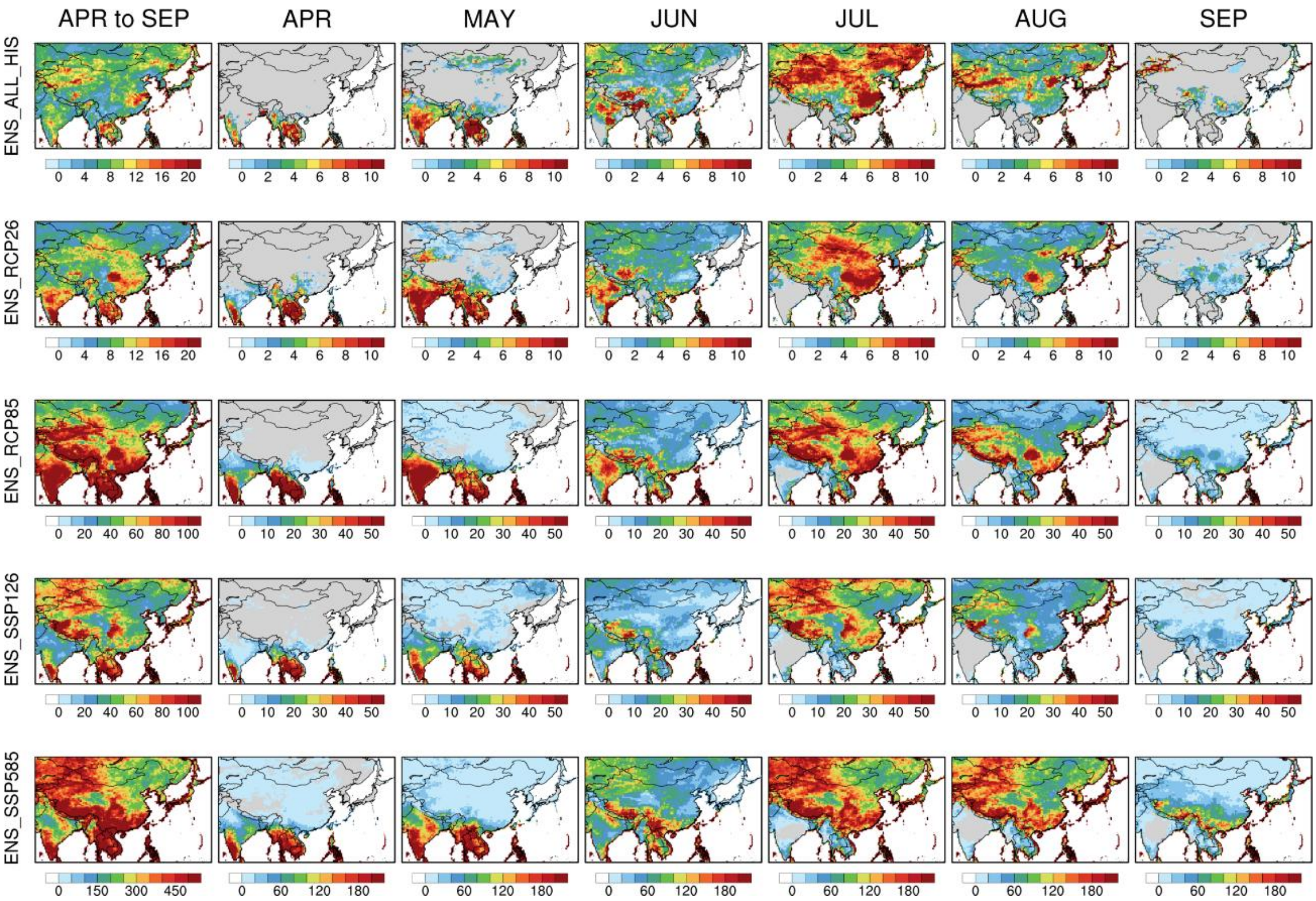

**Figure 9.** Spatial distributions for monthly accumulated intensity of concurrent daytime and nighttime heatwaves (CDNHWs) during the heatwave period in East Asia from April to September for (top to bottom) ENS_ALL_HIS, ENS_RCP26, ENS_RCP85, ENS_SSP126, and ENS_SSP585.

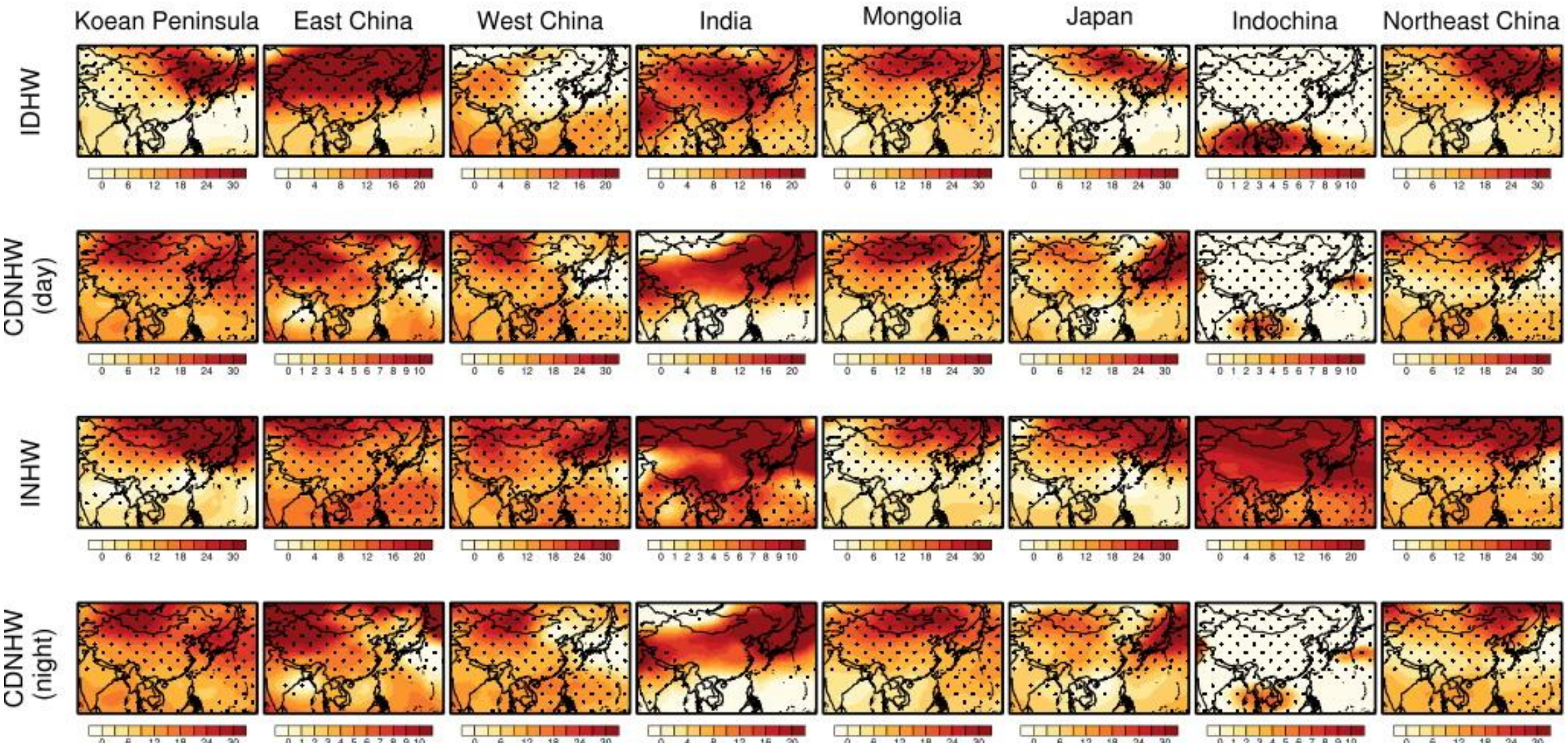

**Figure 10**. Composites of geopotential height at 500hPa anomalies during daytime for IDHWs (top row) and CDNHWs (second row), and during nighttime for INHWs (third row) and CDNHWs (bottom row) over eight sub-regions (Unit: m) in Historical simulation. The grid points where the anomaly is significant at the 95% confidence level based on the Student`s t-test are marked with black dots.

**Table S1.** Twelve GCM-RCM chains and the configurations used in this study.

| GCM-RCM chains | HG2_ CCLM | HG2_ RegCM | HG2_ MM5 | MPI_ WRF | MPI_ CCLM | MPI_ MM5 | GFDL_ WRF | GFDL_ RegCM |
|---|---|---|---|---|---|---|---|---|
| GCM (CMIP phase) | HadGEM2-AO (CMIP5) | | | MPI-ESM-LR (CMIP5) | | | GFDL-ESM2M (CMIP5) | |
| RCM | CCLM | RegCM | SNU-MM5 | WRF | CCLM | SNU-MM5 | WRF | RegCM |
| Forcing Scenarios | Representative Concentration Pathways (RCP2.6 and RCP8.5) | | | | | | | |
| Long /short waves | Ritter and Geleyn | NCAR CCM3 | CCM2 package | CAM | Ritter and Geleyn | CCM2 package | CAM | NCAR CCM3 |
| Explicit moisture scheme | Extended DM | SUBEX | Reisner2 | WSM3 | Extended DM | Reisner2 | WSM3 | SUBEX |
| Land Surface | TERRA-ML | NCAR CLM3.5 | NCAR CLM3/NOAH LSM | Noah | TERRA-ML | NCAR CLM3/NOAH LSM | NOAH LSM | NCAR CLM3.5 |
| PBL | Davis and Turner | Holtslag | YSU | YSU | Davis and Turner | YSU | YSU | Holtslag |
| Cumulus | Tiedtke-mass-flux-convection | MIT-Emanuel | Kain-Fritch | Bats-Miller-Janjic | Tiedtke-mass-flux-convection | Kain-Fritch | Bats-Miller-Janjic | MIT-Emanuel |
| lat x lon | 251x396 | 249x394 | 260x405 | 250x395 | 251x396 | 260x405 | 250x395 | 249x394 |

| UKESM_ WRF | UKESM_ CCLM | UKESM_ GRIMs | UKESM_ RegCM |
|---|---|---|---|
| UK-ESM (CMIP6) | | | |
| WRF | CCLM | GRIMs | RegCM |
| Shared Socioeconomic Pathways (SSP126 and SSP585) | | | |
| CAM | Ritter and Geleyn | Chou et al., 1999/ Chou and Suarez, 1999 | NCAR CCM3 |
| WSM3 | Extended DM | - | SUBEX |
| NOAH LSM | TERRA-ML | NOAH LSM | NCAR CLM4.5 |
| YSU | Davis and Turner | YSU | Holtslag |
| Bats-Miller-Janjic | Tiedtke-mass-flux-convection | Simplified Arakawa-Schubert | MIT-Emanuel |
| 250x395 | 231x376 | 252x401 | 248x393 |

**Table S2.** The first dates when concurrent daytime and nighttime heatwaves occur (FD_CDNHW) in the eight sub-domains (R1~R8).

| | ENS_ALL_HIS | ENS_RCP26 | ENS_RCP85 | ENS_SSP126 | ENS_SSP585 |
|---|---|---|---|---|---|
| R1 | From late July to early August | Late July | From early to mid-July | Mid-July | Mid-June |
| R2 | Mid-July | From late June to early July | From early to mid-June | From mid- to late June | From mid- to late May |
| R3 | Mid-July | Early July | Mid-June | From mid- to late June | Late May |
| R4 | May | From late April to mid-May | April | From mid-April to mid-May | Early April |
| R5 | Early July | Early July | Mid-June | Mid-June | Late May |
| R6 | From late July to early August | Late July | Mid-July | Mid-July | Mid-June |
| R7 | From late April to late May | From mid-April to mid-May | From early to mid-April | April | Early April |
| R8 | Mid-July | Mid-July | Late June | From mid- to late June | From late May to early June |

**Table S3.** The last dates when concurrent daytime and nighttime heatwaves occur (LD_CDNHW) in the eight sub-domains (R1~R8).

| | ENS_ALL_HIS | ENS_RCP26 | ENS_RCP85 | ENS_SSP126 | ENS_SSP585 |
|---|---|---|---|---|---|
| R1 | From early to mid-August | From early to mid-August | From mid-August to early September | From late August to mid-September | From mid- to late September |
| R2 | From mid-July to early August | From late-July to early August | From mid-August to early September | From mid- to late August | Mid-September |
| R3 | From mid-July to early August | From mid-July to early August | Late August | From mid- to late August | Mid-September |
| R4 | From mid-May to early June | From late May to early June | From early June to mid-August | From late May to mid-June | From mid-June to mid-July |
| R5 | Late July | Late July | Mid-August | Mid-August | Early September |
| R6 | Mid-August | Mid- to late August | From late August to mid-September | From early to mid-September | Late September |
| R7 | From early May to early June | From mid-May to mid-June | From late June to early September | From early June to late August | September |
| R8 | Late July | Late July | Early August | Mid-August | Early September |

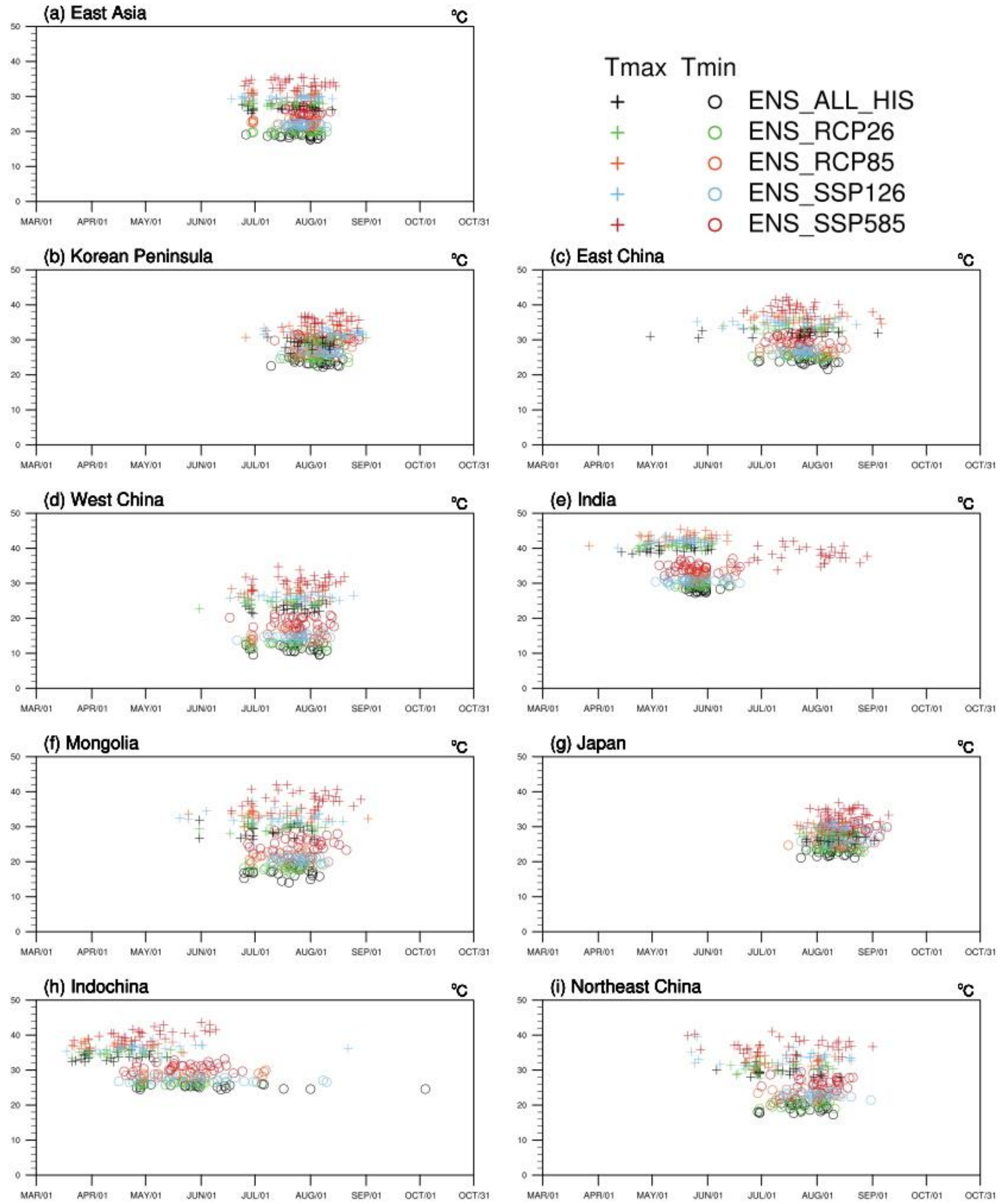

**Figure S1.** Scatter plots of annual maximum values for daily maximum and minimum temperatures (Tmax and Tmin) and the dates when those appear (a) on the Korean Peninsula, (b) in East China, (c) in West China, (d) in India, (e) in Mongolia, (f) in Japan, (g) in Indochina, and (h) in Northeast China. Black, green, orange, blue, and red open circles (crosses) denote the Tmax (Tmin) of each year during the Historical reference period (1981-2005) and in the future period (2071-2100).

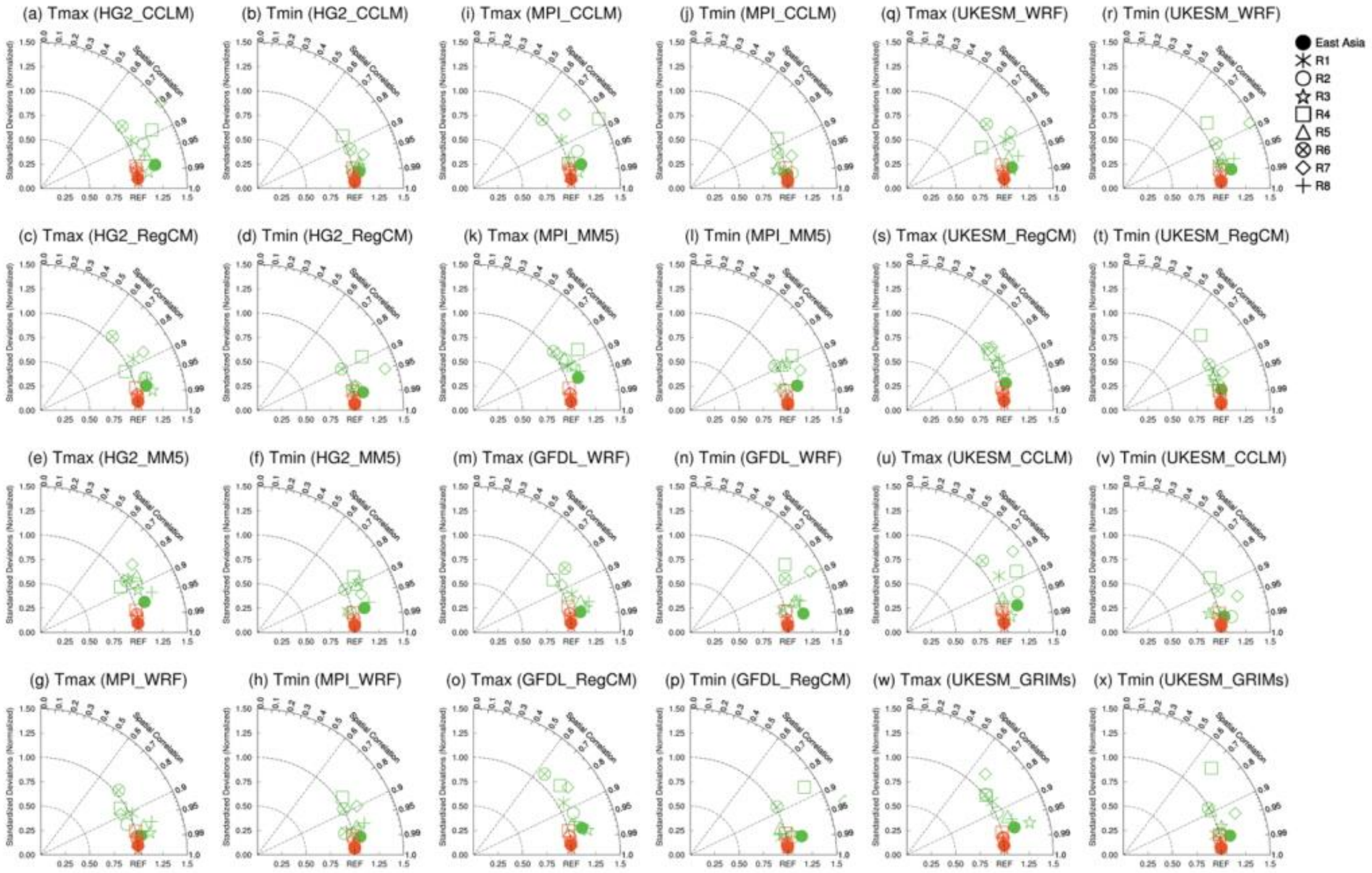

**Figure S2.** Taylor diagrams of the 25-year (1981-2005) mean daily maximum (Tmax) and minimum temperatures (Tmin) during the heatwave period in East Asia derived from twelve GCM-RCM chains. The radial axes show the temporal standard deviation of the GCM-RCM chain normalized with ERA5, and the arcs denote the spatial correlation coefficients between the GCM-RCM chain and ERA5. The REF point (1.0) denotes where the GCM-RCM chain exactly agrees with ERA5. In the legend, _ORG and _VS, respectively, indicate non-bias-corrected and bias-corrected Tmax or Tmin.

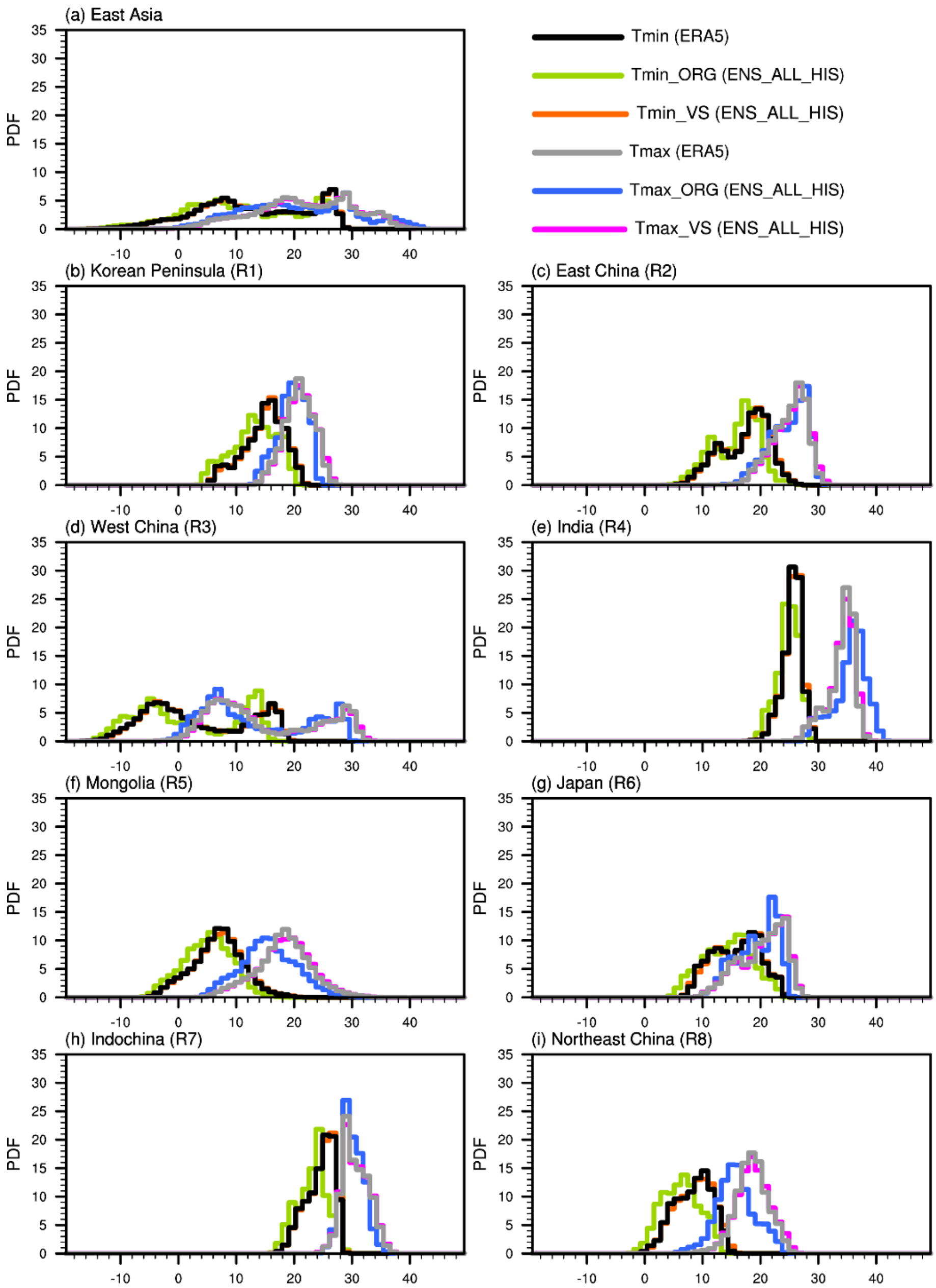

**Figure S3.** Probability distribution functions for mean daily maximum temperature (Tmax) and mean daily minimum temperature (Tmin) during the heatwave period of East Asia (a) in East Asia, (b) in South Korea, (c) in East China, (d) in West China, (e) in India, (f) in Mongolia, (g) in Japan, (h) in Indochina, and (i) in Northeast China (Unit: ºC). Black, green, and orange lines, respectively, denote Tmax derived from ERA5, non-bias corrected Tmax (Tmax_ORG), and bias-corrected Tmax (Tmax_VS), both derived from ENS_ALL_HIS. Grey, blue, and pink lines, respectively, denote Tmin derived from ERA5, non-bias corrected Tmin (Tmin_ORG), and bias-corrected Tmin (Tmin_VS), both derived from ENS_ALL_HIS.

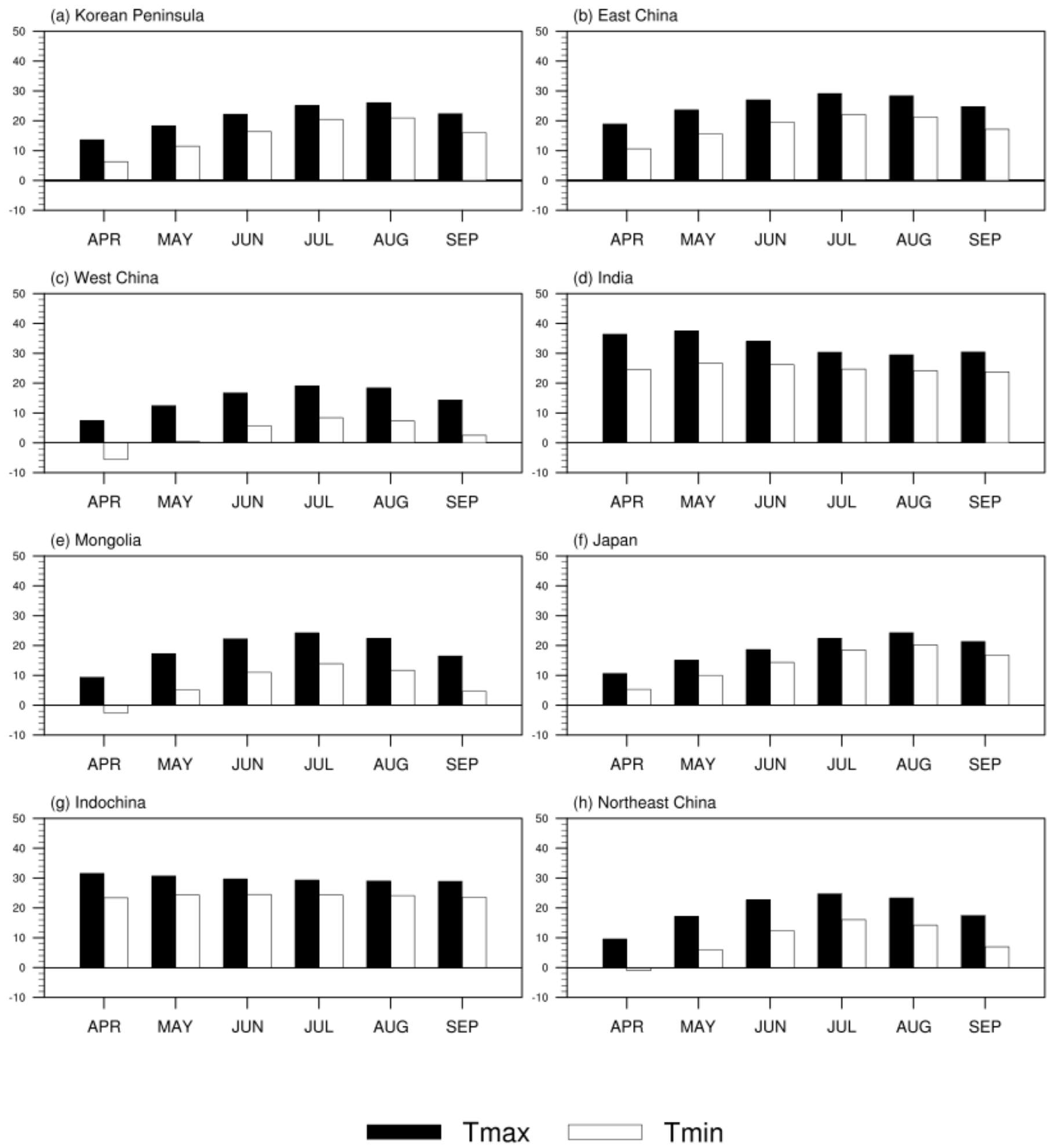

**Figure S4.** Climatological monthly daily maximum temperatures (Tmax) and daily minimum temperatures (Tmin) (a) over the Korean Peninsula, (b) in East China, (c) in West China, (d) in India, (e) in Mongolia, (f) in Japan, (g) in Indochina, and (h) in Northeast China (Unit: ºC). Black and white bars denote Tmax and Tmin, respectively, derived from ENS_ALL_HIS.

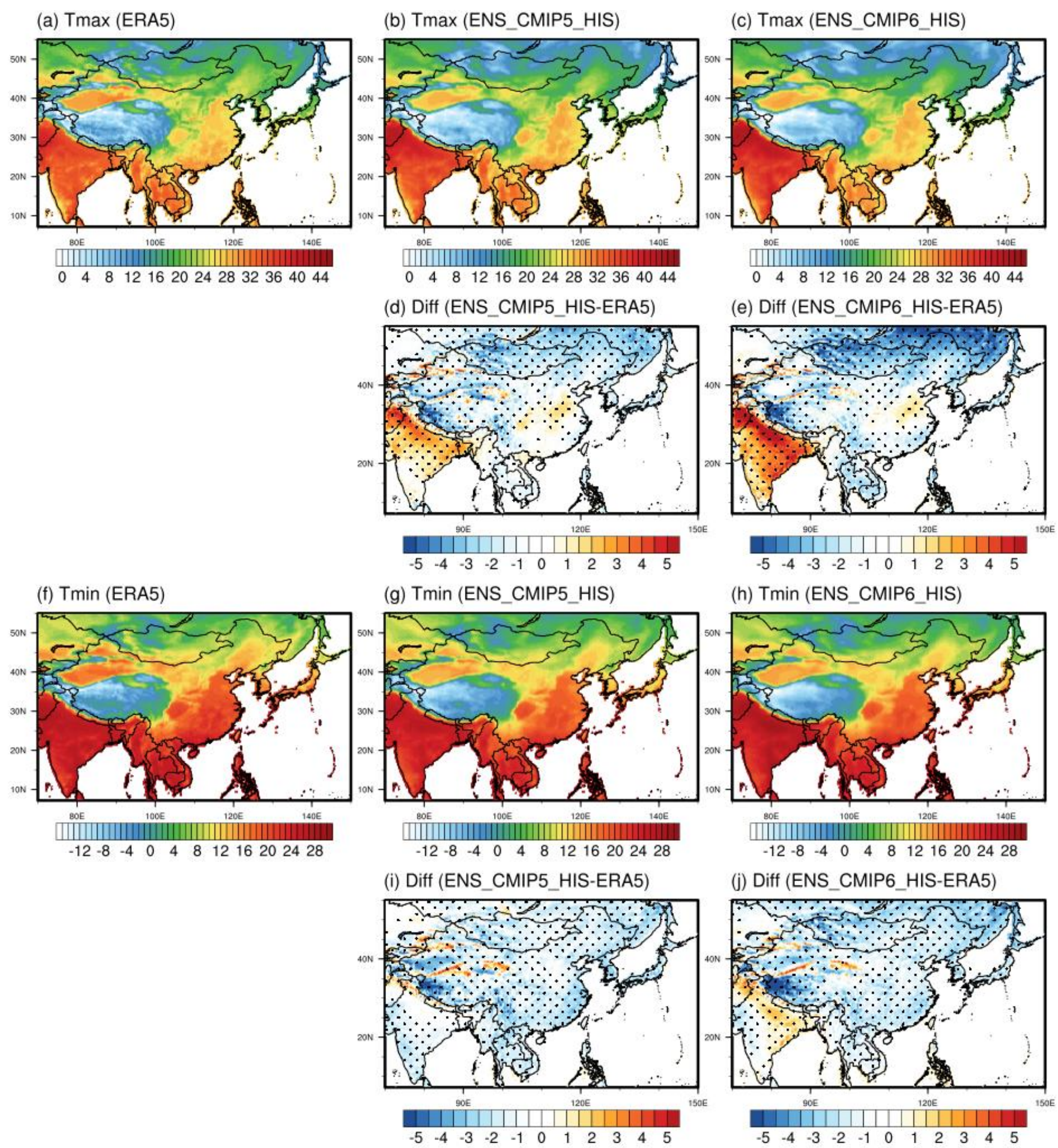

**Figure S5.** Spatial distributions of the mean daily maximum and minimum temperatures during the heatwave period from April to September in East Asia derived from ERA5, ENS_CMIP5_HIS, and ENS_CMIP6_HIS, as well as the bias, compared with ERA5 (Unit: ºC). The grid points where the difference is significant at the 95% confidence level based on the Student`s t-test are marked with black dots.

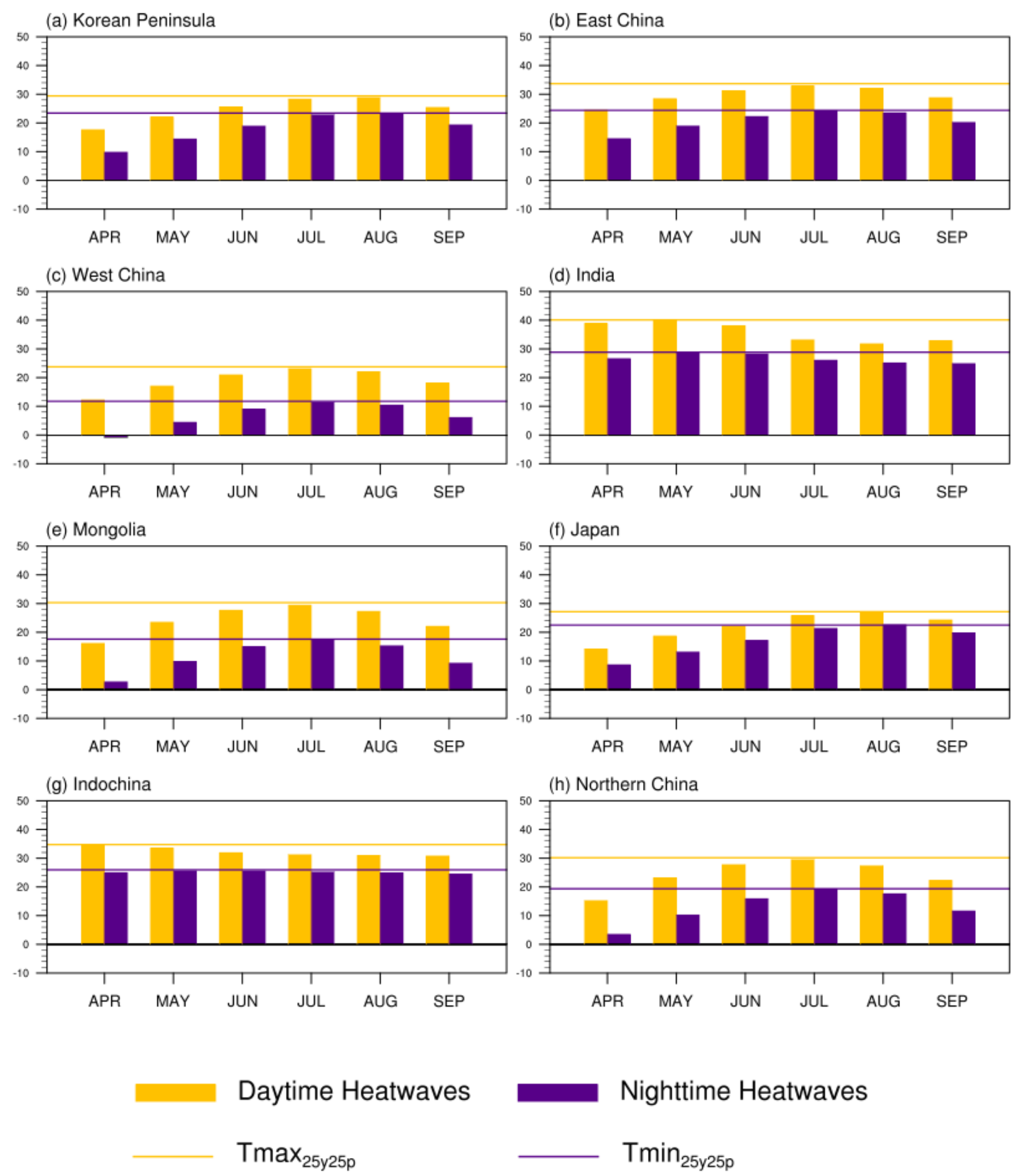

**Figure S6.** Monthly thresholds for daytime and nighttime heatwaves over (a) the Korean Peninsula, (b) East China, (c) West China, (d) India, (e) Mongolia, (f) Japan, (g) Indochina, and (h) Northeast China (Unit: ºC). Yellow and purple bars denote thresholds of daytime and nighttime heatwaves, respectively.

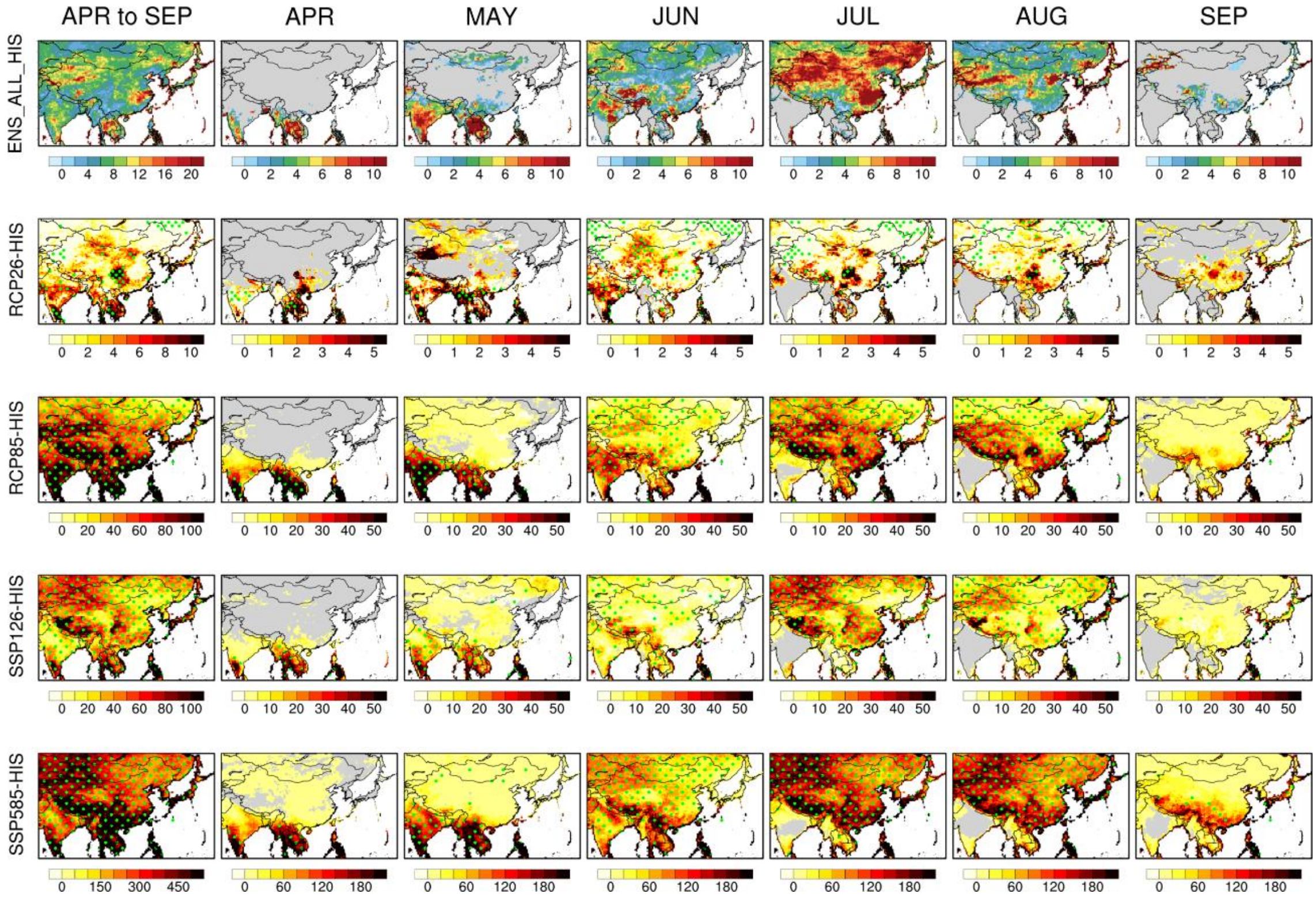

**Figure S7.** Spatial distributions of monthly accumulated intensity for concurrent daytime and nighttime heatwaves (CDNHWs) during the heatwave period of East Asia from April to September derived from (top) ENS_ALL_HIS, and the future changes derived from (second from the top to the bottom) ENS_RCP26, ENS_RCP85, ENS_SSP126, and ENS_SSP585. The grid points where the difference is significant at the 95% confidence level based on the Student`s t-test are marked with green dots.

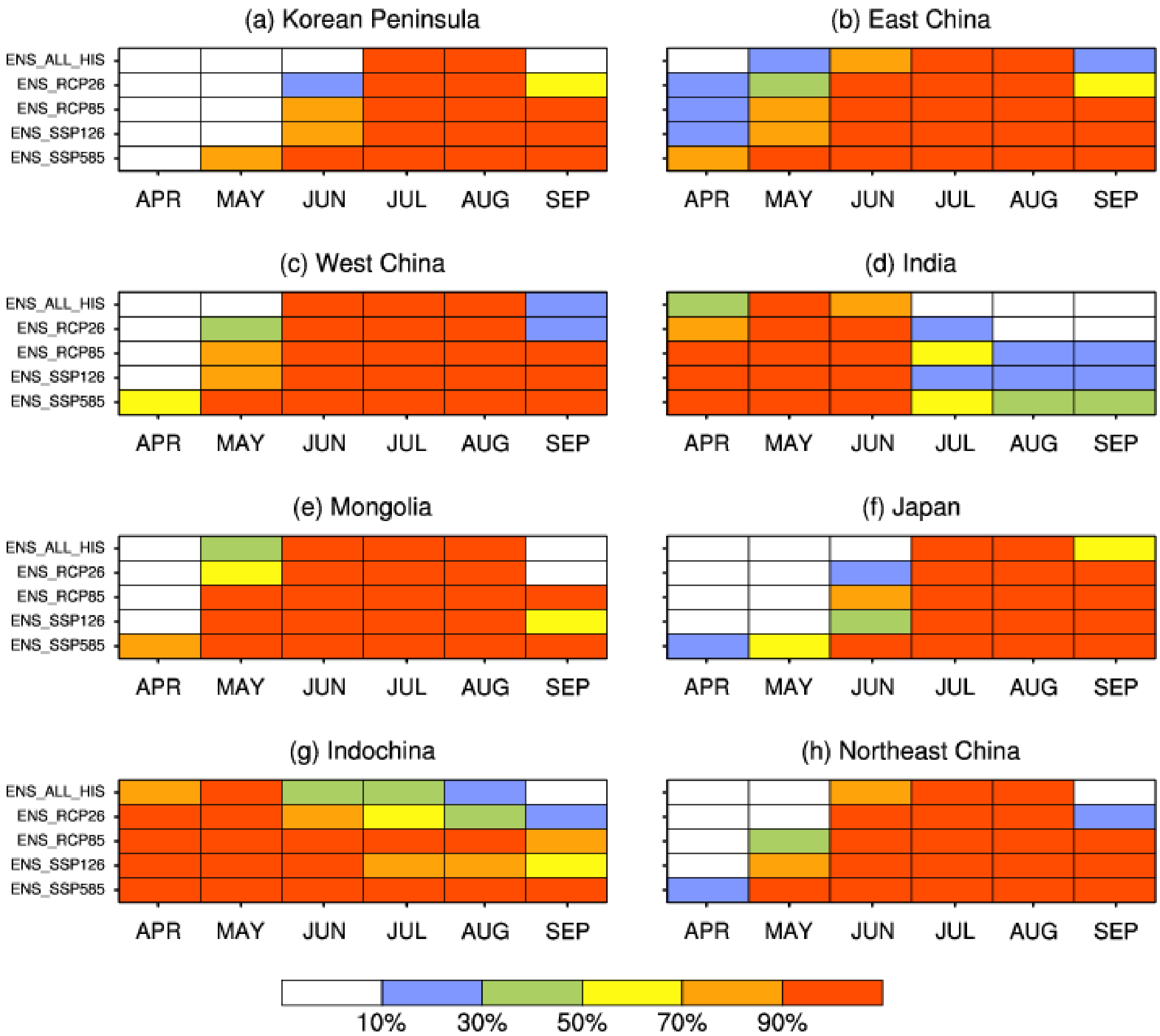

**Figure S8.** The spatial extent of a concurrent daytime and nighttime heatwave (CDNHW), which is calculated as a percentage of CDNHW area to the total area of (a) the Korean Peninsula, (b) East China, (c) West China, (d) India, (e) Mongolia, (f) Japan, (g) Indochina, and (h) Northeast China derived from (top row to bottom row of each chart) ENS_ALL_HIS, ENS_ RCP26, ENS_RCP85, ENS_SSP126, and ENS_SSP585 (Unit: %).

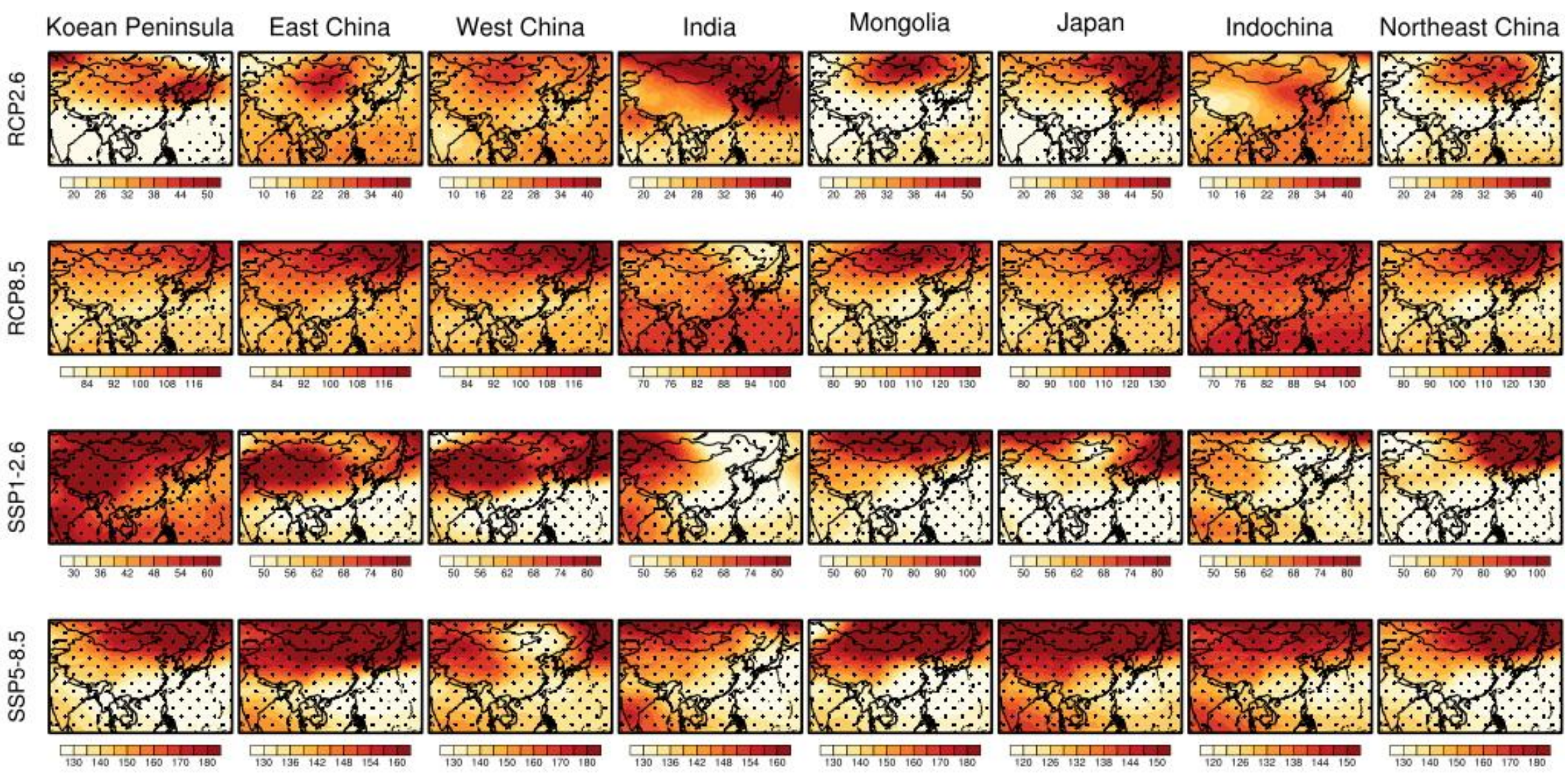

**Figure S9**. Composites of geopotential height at 500hPa anomalies during daytime for CDNHW over eight sub-regions under (top to bottom) Historical, RCP2.6, RCP8.5, SSP1-2.6, and SSP5-8.5 scenarios (Unit: m). The period, 1981-2005 was used as reference period for calculating climatology.

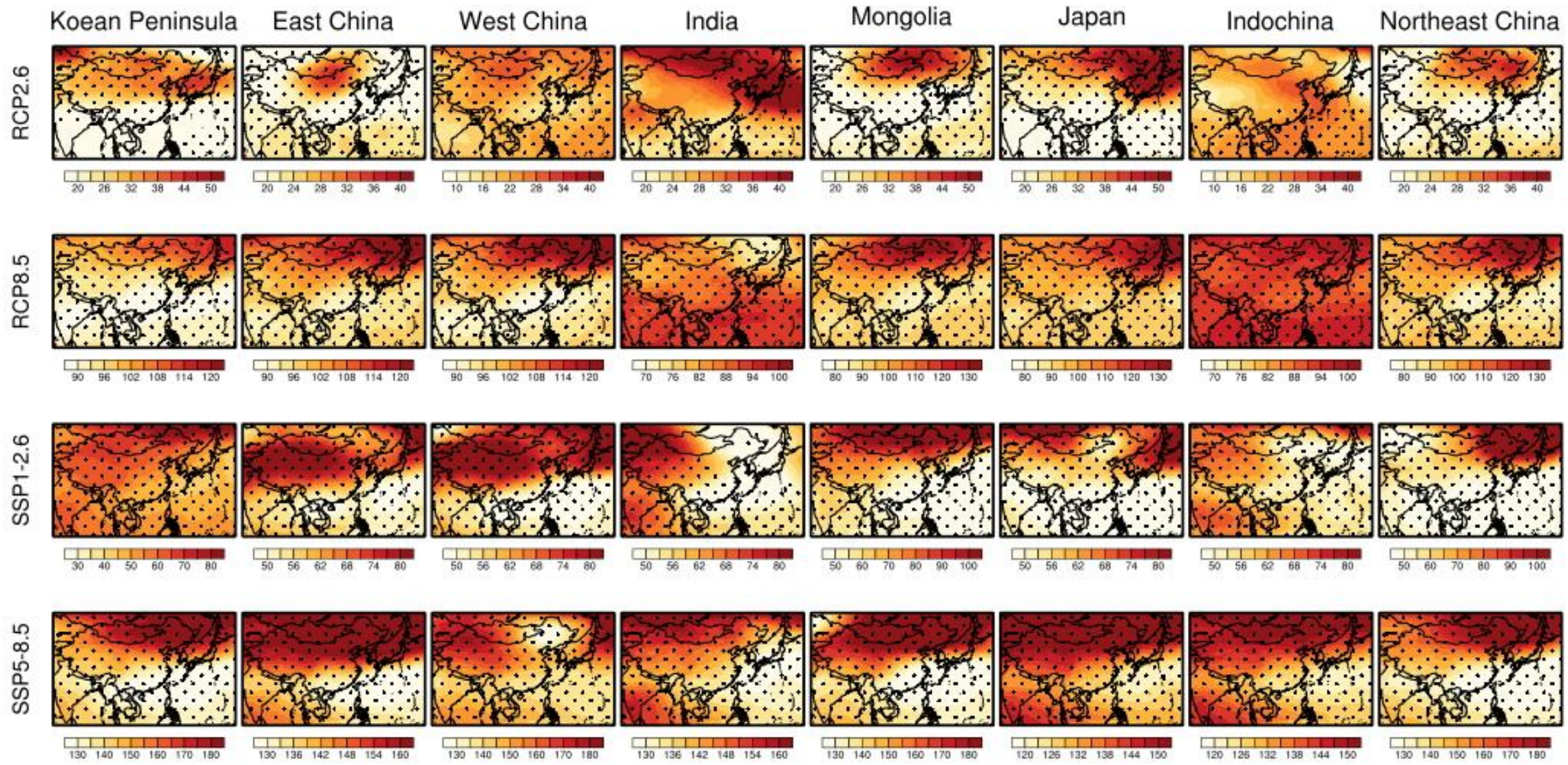

**Figure S10.** Composites of geopotential height at 500hPa anomalies during nighttime for CDNHW over eight sub-regions under (top to bottom) Historical, RCP2.6, RCP8.5, SSP1-2.6, and SSP5-8.5 scenarios (Unit: m). The period, 1981-2005 was used as reference period for calculating climatology.

Supporting information for

# Concurrent daytime and nighttime heatwaves in the late 21st century over the CORDEX-East Asia phase 2 domain using Multi-GCM and Multi-RCM Chains

**Young-Hyun Kim[1], Joong-Bae Ahn[1*], Myoung-Seok Suh[2,] Dong-Hyun Cha[3], Eun-Chul Chang[2], Seung-Ki Min[3], Young-Hwa Byun[5] and Jin-Uk Kim[5]**

[1]Department of Atmospheric Sciences, Pusan National University, Busan, Korea

[2]Department of Atmospheric Science, Kongju National University, Gongju, South Korea

[3]Department of Urban and Environmental Engineering, Ulsan National Institute of Science and Technology, Ulsan, South Korea

[4]Division of Environmental Science and Engineering, Pohang University of Science and Technology, Pohang, South Korea

[5]Climate Change Research Team, National Institute of Meteorological Sciences

**Table S1.** Twelve GCM-RCM chains and the configurations used in this study.

| GCM-RCM chains | HG2_ CCLM | HG2_ RegCM | HG2_ MM5 | MPI_ WRF | MPI_ CCLM | MPI_ MM5 | GFDL_ WRF | GFDL_ RegCM |
|---|---|---|---|---|---|---|---|---|
| GCM (CMIP phase) | HadGEM2-AO (CMIP5) | | | MPI-ESM-LR (CMIP5) | | | GFDL-ESM2M (CMIP5) | |
| RCM | CCLM | RegCM | SNU-MM5 | WRF | CCLM | SNU-MM5 | WRF | RegCM |
| Forcing Scenarios | Representative Concentration Pathways (RCP2.6 and RCP8.5) | | | | | | | |
| Long /short waves | Ritter and Geleyn | NCAR CCM3 | CCM2 package | CAM | Ritter and Geleyn | CCM2 package | CAM | NCAR CCM3 |
| Explicit moisture scheme | Extended DM | SUBEX | Reisner2 | WSM3 | Extended DM | Reisner2 | WSM3 | SUBEX |
| Land Surface | TERRA-ML | NCAR CLM3.5 | NCAR CLM3/NOAH LSM | Noah | TERRA-ML | NCAR CLM3/NOAH LSM | NOAH LSM | NCAR CLM3.5 |
| PBL | Davis and Turner | Holtslag | YSU | YSU | Davis and Turner | YSU | YSU | Holtslag |
| Cumulus | Tiedtke-mass-flux-convection | MIT-Emanuel | Kain-Fritch | Bats-Miller-Janjic | Tiedtke-mass-flux-convection | Kain-Fritch | Bats-Miller-Janjic | MIT-Emanuel |
| lat x lon | 251x396 | 251x396 | 260x405 | 250x395 | 251x396 | 260x405 | 250x395 | 251x396 |

| UKESM_ WRF | UKESM_ CCLM | UKESM_ GRIMs | UKESM_ RegCM |
|---|---|---|---|
| UK-ESM (CMIP6) | | | |
| WRF | CCLM | GRIMs | RegCM |
| Shared Socioeconomic Pathways (SSP126 and SSP585) | | | |
| CAM | Ritter and Geleyn | Chou et al., 1999/ Chou and Suarez, 1999 | NCAR CCM3 |
| WSM3 | Extended DM | - | SUBEX |
| NOAH LSM | TERRA-ML | NOAH LSM | NCAR CLM4.5 |
| YSU | Davis and Turner | YSU | Holtslag |
| Bats-Miller-Janjic | Tiedtke-mass-flux-convection | Simplified Arakawa-Schubert | MIT-Emanuel |
| 250x395 | 251x396 | 252x401 | 251x396 |

**Table S2.** The first dates when concurrent daytime and nighttime heatwaves occur (FD_CDNHW) in the eight sub-domains (R1~R8).

| | ENS_ALL_HIS | ENS_RCP26 | ENS_RCP85 | ENS_SSP126 | ENS_SSP585 |
|---|---|---|---|---|---|
| R1 | From late July to early August | Late July | From early to mid-July | Mid-July | Mid-June |
| R2 | Mid-July | From late June to early July | From early to mid-June | Mid-June | From early to mid-May |
| R3 | Mid-July | Early July | Mid-June | From mid- to late June | Late May |
| R4 | May | From late April to mid-May | April | From mid-April to mid-May | Early April |
| R5 | Early July | Early July | Mid-June | Mid-June | Late May |
| R6 | From late July to early August | Late July | Mid-July | Mid-July | Mid-June |
| R7 | From late April to late May | From mid-April to mid-May | From early to mid-April | April | Early April |
| R8 | Mid-July | Mid-July | Late June | From mid- to late June | From late May to early June |

**Table S3.** The last dates when concurrent daytime and nighttime heatwaves occur (LD_CDNHW) the eight sub-domains (R1~R8).

| | ENS_ALL_HIS | ENS_RCP26 | ENS_RCP85 | ENS_SSP126 | ENS_SSP585 |
|---|---|---|---|---|---|
| R1 | From early to mid-August | From early to mid-August | From mid-August to early September | From late August to mid-September | From mid- to late September |
| R2 | From mid-July to early August | From late-July to early August | From mid-August to early September | From mid- to late August | Mid-September |
| R3 | From mid-July to early August | From mid-July to early August | Late August | From mid- to late August | Mid-September |
| R4 | From mid-May to early June | From late May to early June | From early June to mid-August | From late May to mid-June | From mid-June to mid-July |
| R5 | Late July | Late July | Mid-August | Mid-August | Early September |
| R6 | Mid-August | Mid- to late August | From late August to mid-September | From early to mid-September | Late September |
| R7 | From early May to early June | From mid-May to mid-June | From late June to early September | From early June to late August | September |
| R8 | Late July | Late July | Early August | Mid-August | Early September |

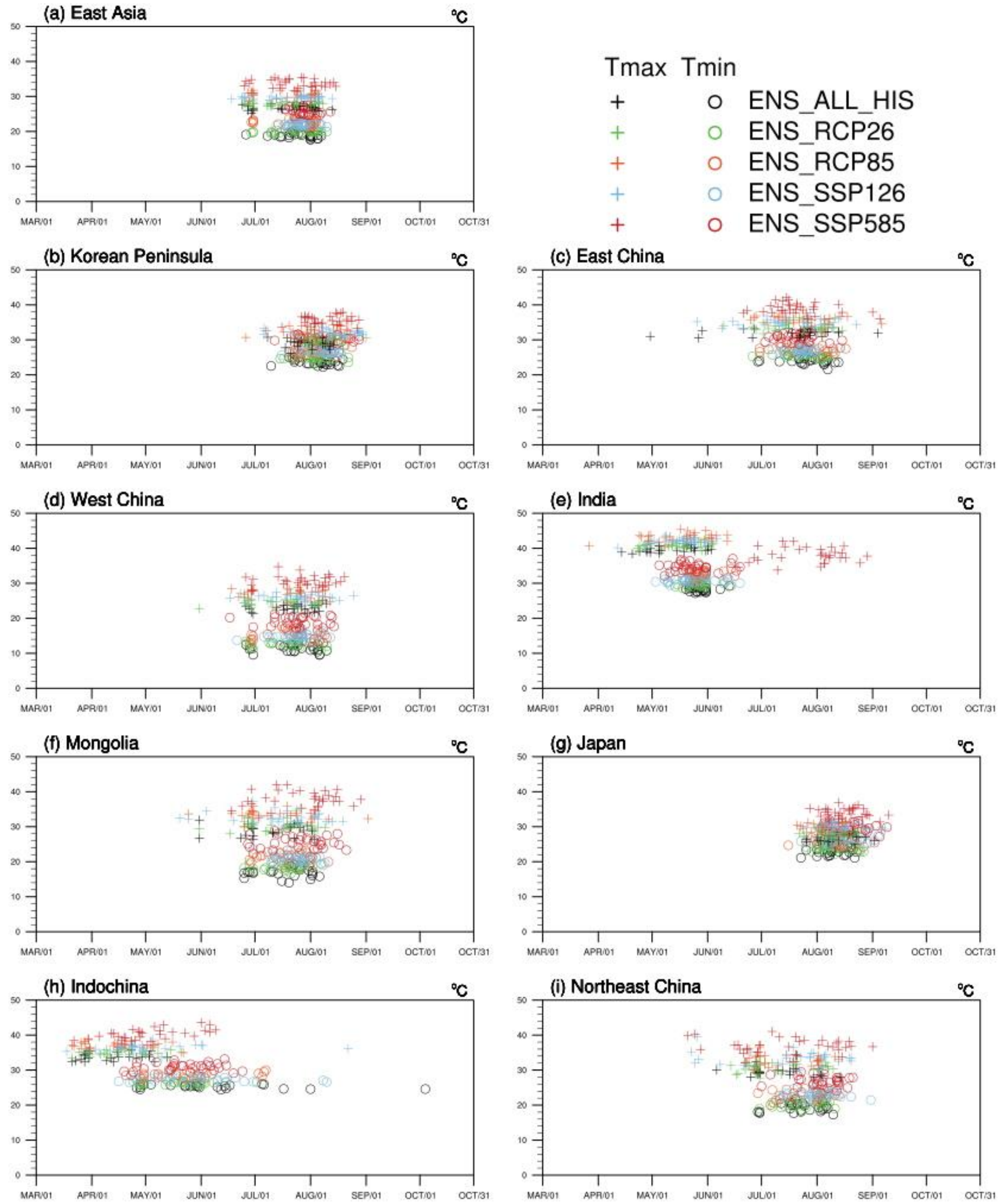

**Figure S1.** Scatter plots of annual maximum values for daily maximum and minimum temperatures (Tmax and Tmin) and the dates when those appear (a) on the Korean Peninsula, (b) in East China, (c) in West China, (d) in India, (e) in Mongolia, (f) in Japan, (g) in Indochina, and (h) in Northeast China. Black, green, orange, blue, and red open circles (crosses) denote the Tmax (Tmin) of each year during the Historical reference period (1981-2005) and in the future period (2071-2100).

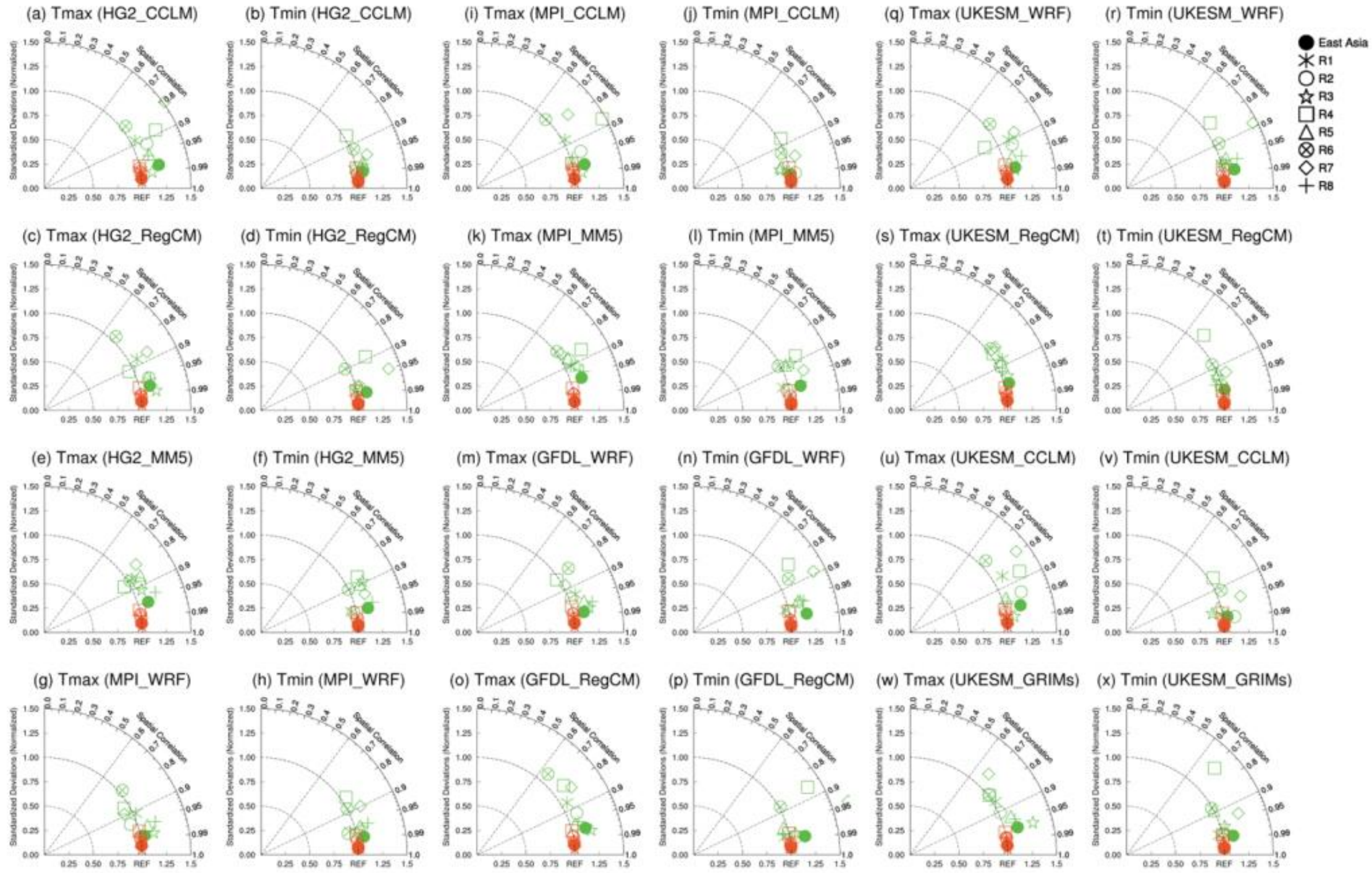

**Figure S2.** Taylor diagrams of the 25-year (1981-2005) mean daily maximum (Tmax) and minimum temperatures (Tmin) during the heatwave period in East Asia derived from twelve GCM-RCM chains. The radial axes show the temporal standard deviation of the GCM-RCM chain normalized with ERA5, and the arcs denote the spatial correlation coefficients between the GCM-RCM chain and ERA5. The REF point (1.0) denotes where the GCM-RCM chain exactly agrees with ERA5. In the legend, _ORG and _VS, respectively, indicate non-bias-corrected and bias-corrected Tmax or Tmin.

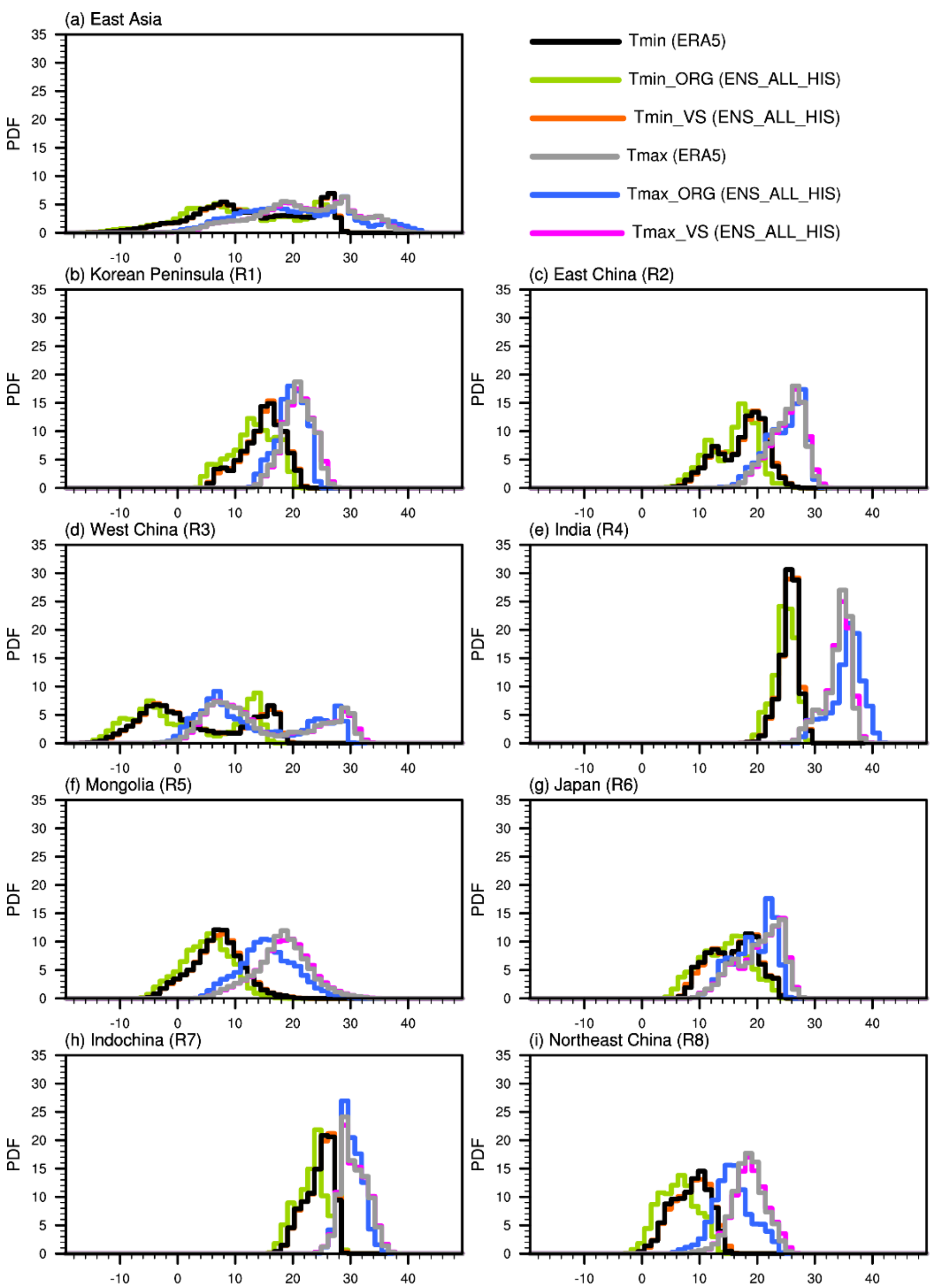

**Figure S3.** Probability distribution functions for mean daily maximum temperature (Tmax) and mean daily minimum temperature (Tmin) during the heatwave period of East Asia (a) in East Asia, (b) in South Korea, (c) in East China, (d) in West China, (e) in India, (f) in Mongolia, (g) in Japan, (h) in Indochina, and (i) in Northeast China (Unit: ºC). Black, green, and orange lines, respectively, denote Tmax derived from ERA5, non-bias corrected Tmax (Tmax_ORG), and bias-corrected Tmax (Tmax_VS), both derived from ENS_ALL_HIS. Grey, blue, and pink lines, respectively, denote Tmin derived from ERA5, non-bias corrected Tmin (Tmin_ORG), and bias-corrected Tmin (Tmin_VS), both derived from ENS_ALL_HIS.

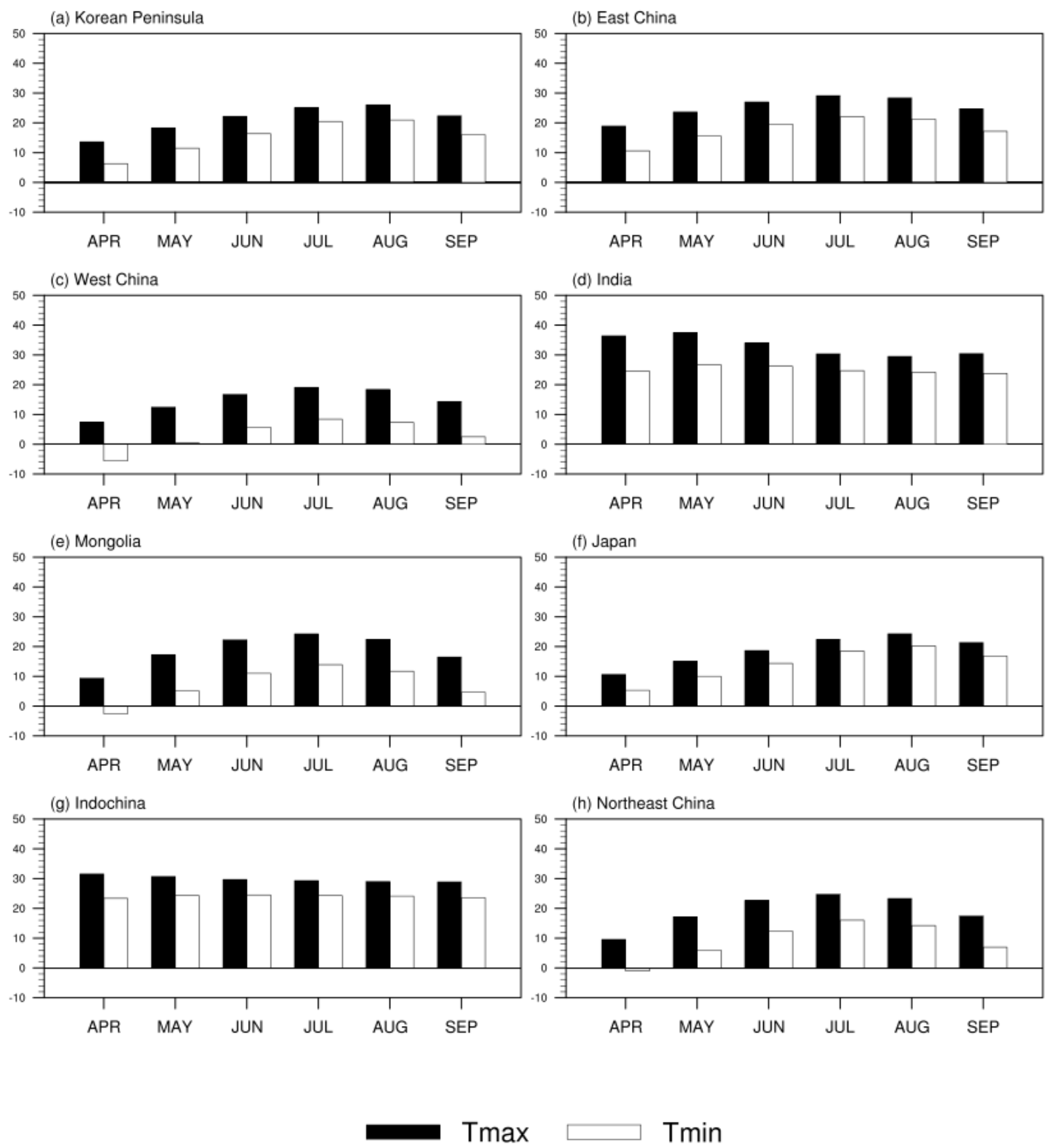

**Figure S4.** Climatological monthly daily maximum temperatures (Tmax) and daily minimum temperatures (Tmin) (a) over the Korean Peninsula, (b) in East China, (c) in West China, (d) in India, (e) in Mongolia, (f) in Japan, (g) in Indochina, and (h) in Northeast China (Unit: ºC). Black and white bars denote Tmax and Tmin, respectively, derived from ENS_ALL_HIS.

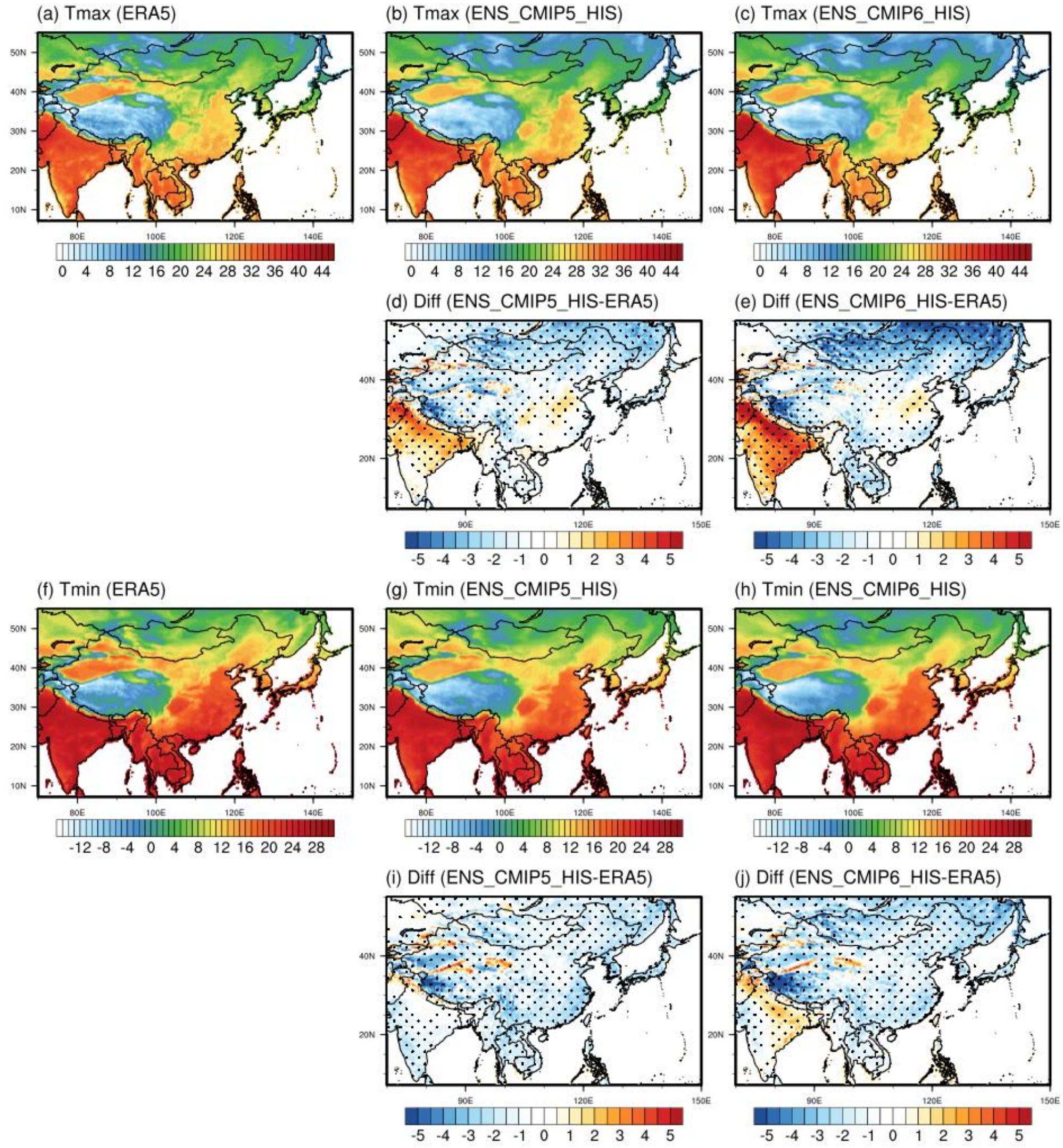

**Figure S5.** Spatial distributions of the mean daily maximum and minimum temperatures during the heatwave period from April to September in East Asia derived from ERA5, ENS_CMIP5_HIS, and ENS_CMIP6_HIS, as well as the bias, compared with ERA5 (Unit: ºC). The grid points where the difference is significant at the 95% confidence level based on the Student`s t-test are marked with black dots.

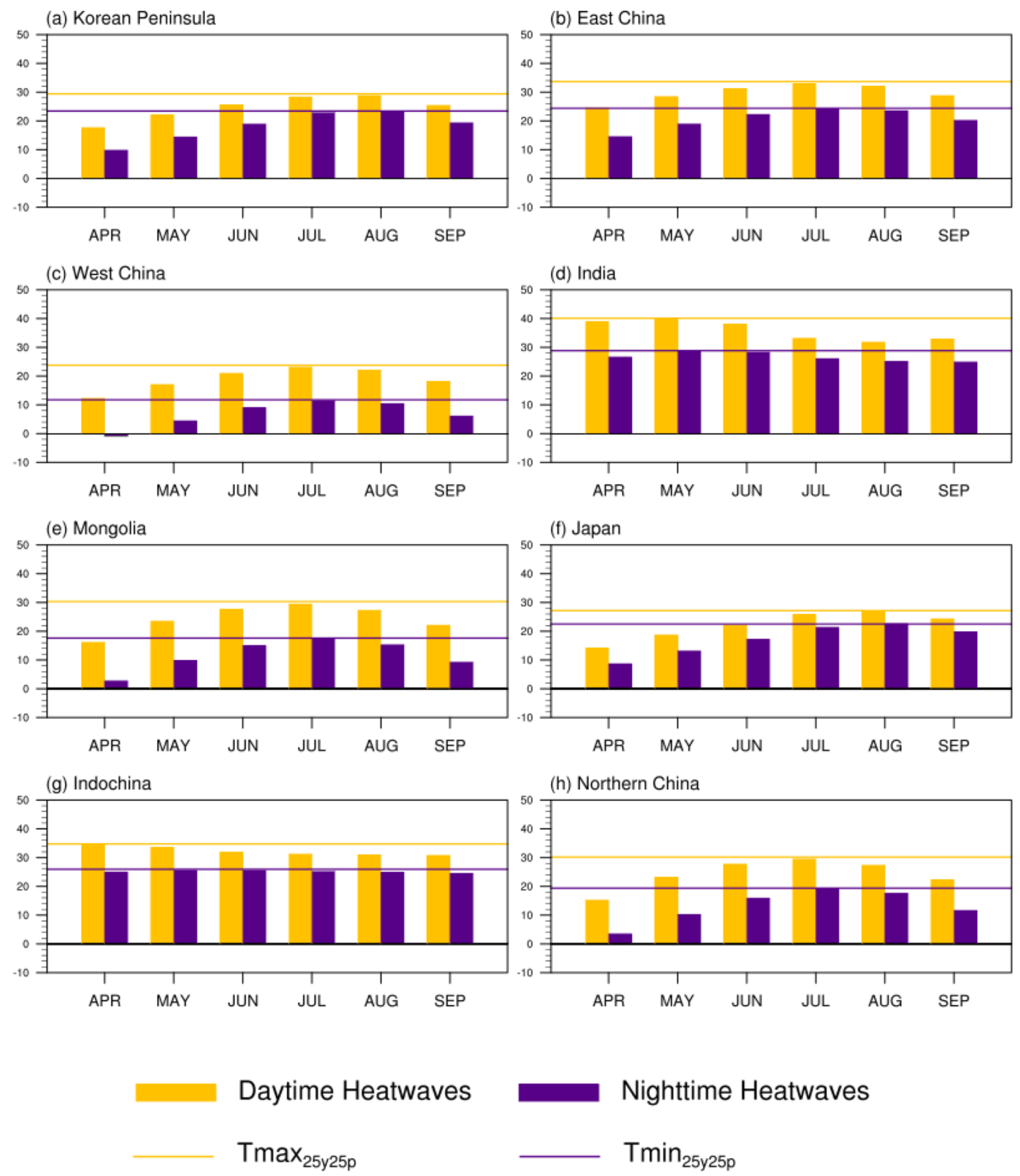

**Figure S6.** Monthly thresholds for daytime and nighttime heatwaves over (a) the Korean Peninsula, (b) East China, (c) West China, (d) India, (e) Mongolia, (f) Japan, (g) Indochina, and (h) Northeast China (Unit: ºC). Yellow and purple bars denote thresholds of daytime and nighttime heatwaves, respectively.

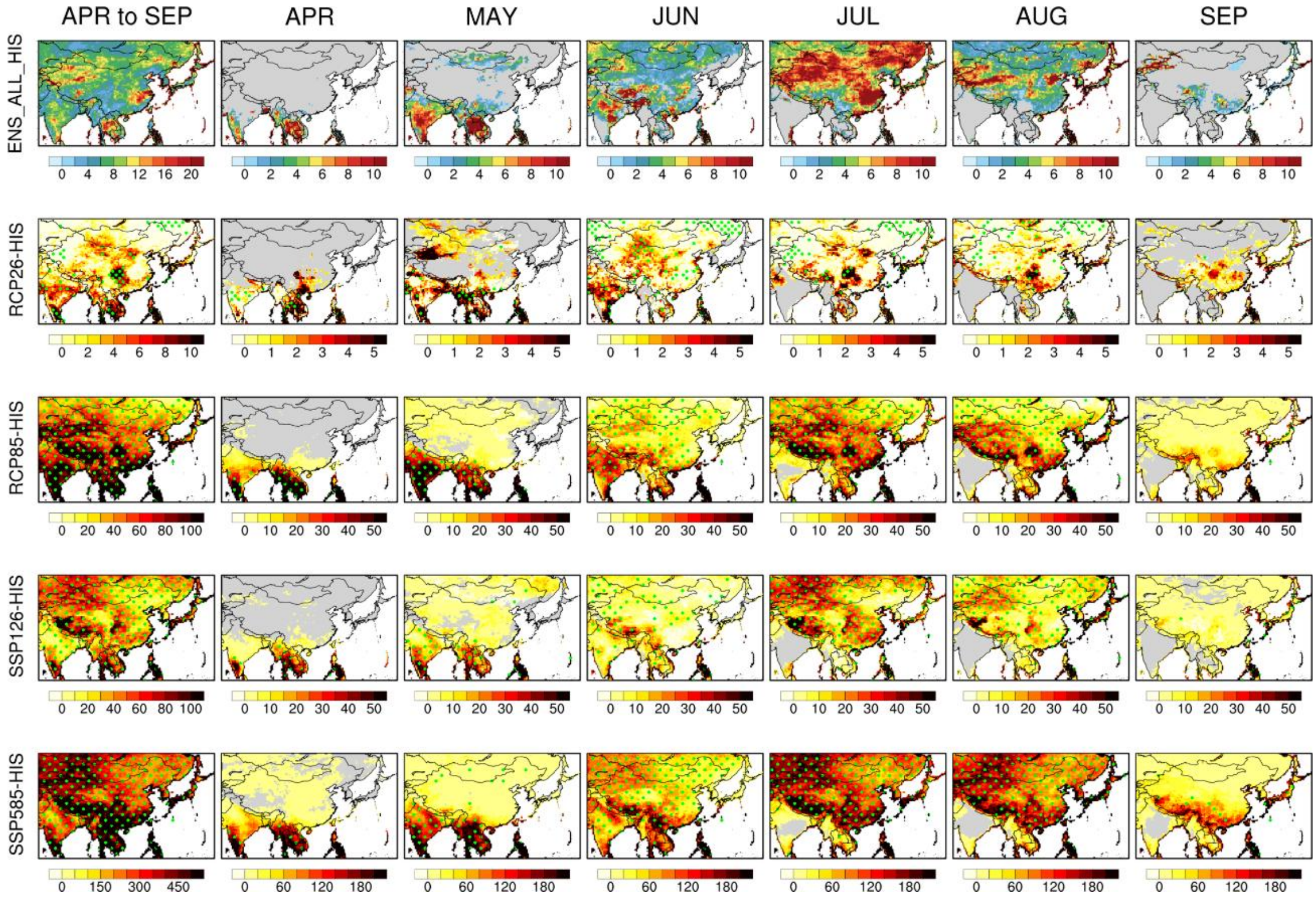

**Figure S7.** Spatial distributions of monthly accumulated intensity for concurrent daytime and nighttime heatwaves (CDNHWs) during the heatwave period of East Asia from April to September derived from (top) ENS_ALL_HIS, and the future changes derived from (second from the top to the bottom) ENS_RCP26, ENS_RCP85, ENS_SSP126, and ENS_SSP585. The grid points where the difference is significant at the 95% confidence level based on the Student`s t-test are marked with green dots.

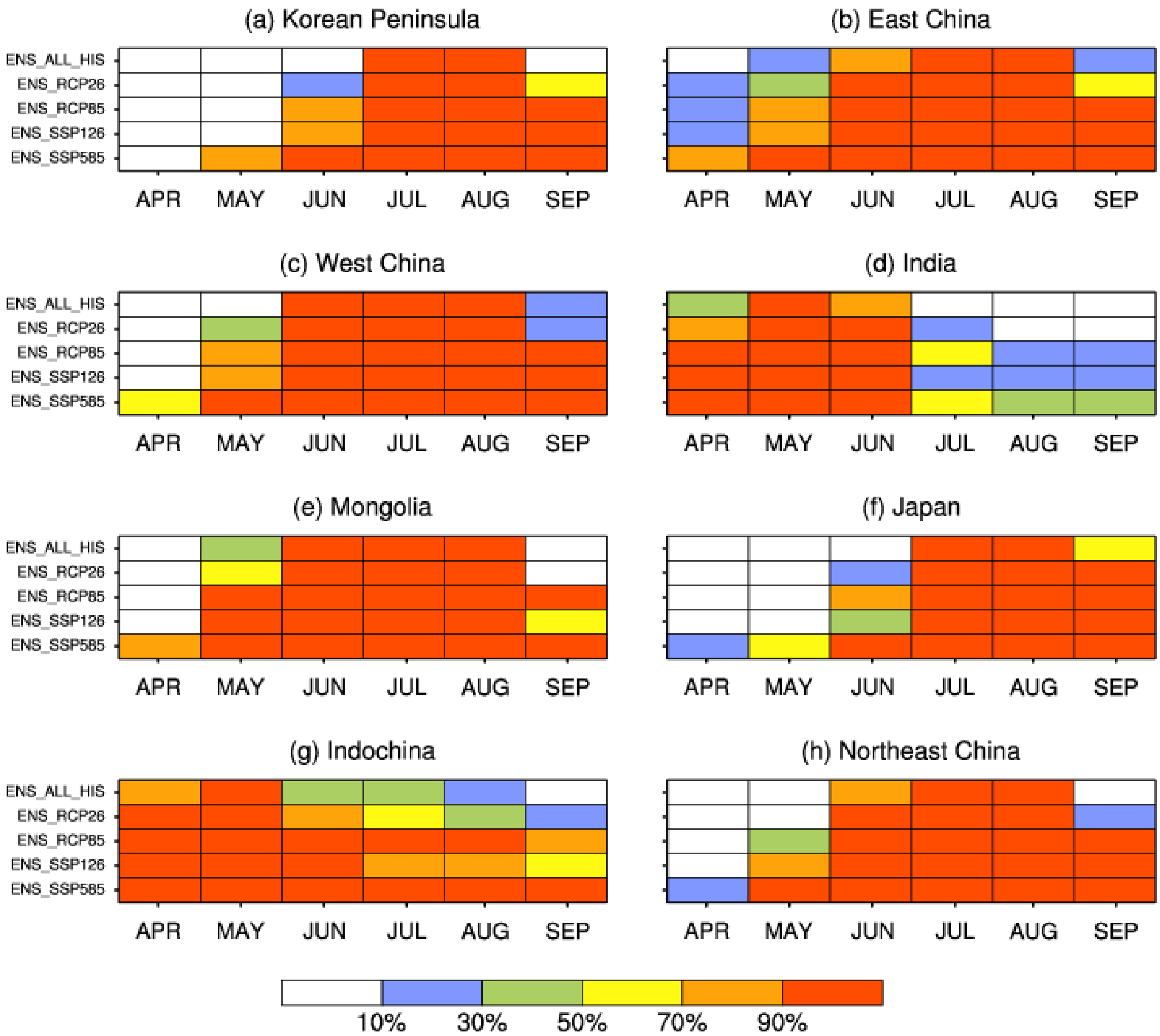

**Figure S8.** The spatial extent of a concurrent daytime and nighttime heatwave (CDNHW), which is calculated as a percentage of CDNHW area to the total area of (a) the Korean Peninsula, (b) East China, (c) West China, (d) India, (e) Mongolia, (f) Japan, (g) Indochina, and (h) Northeast China derived from (top row to bottom row of each chart) ENS_ALL_HIS, ENS_ RCP26, ENS_RCP85, ENS_SSP126, and ENS_SSP585 (Unit: %).

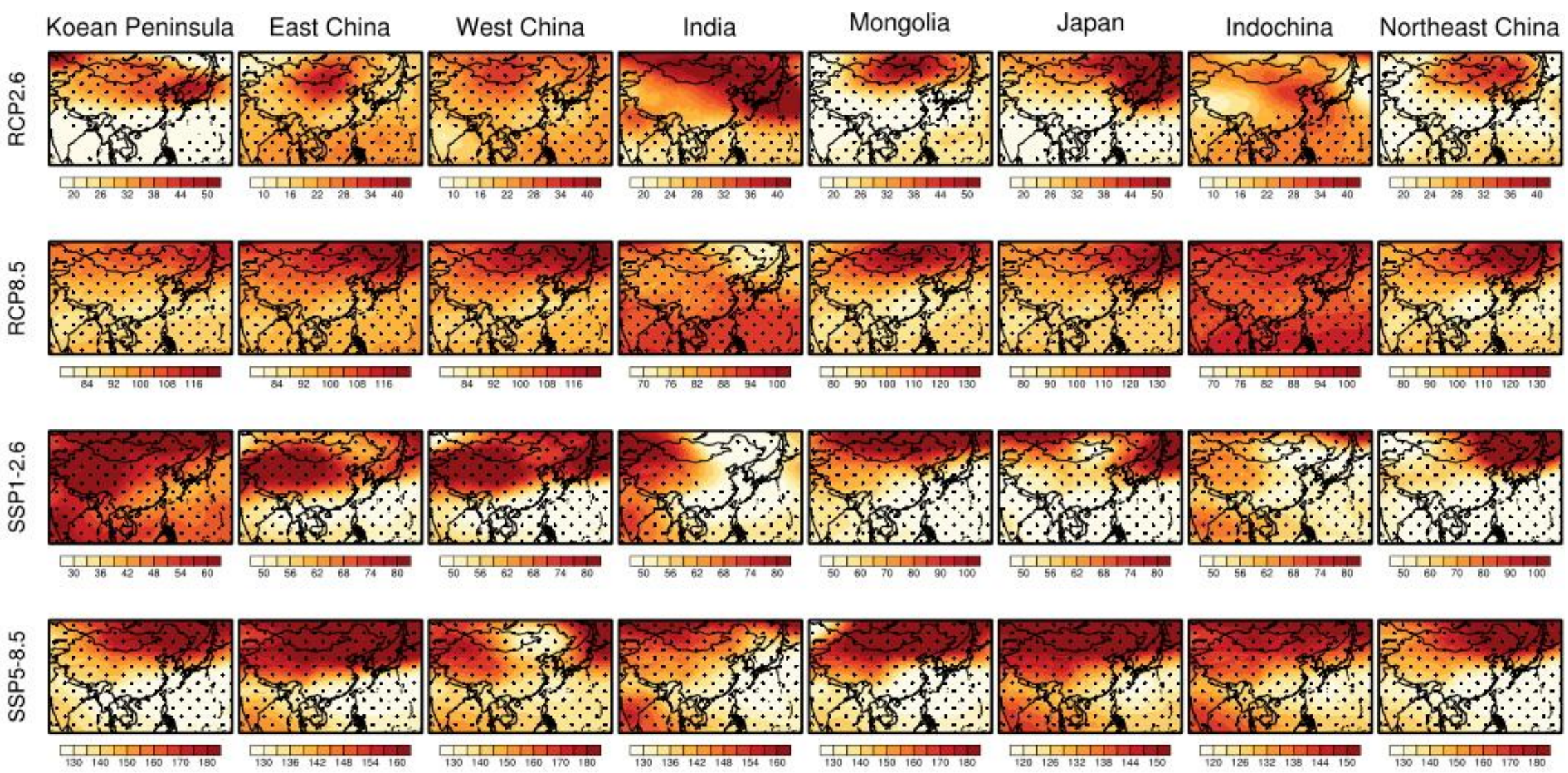

**Figure S9**. Composites of geopotential height at 500hPa anomalies during daytime for CDNHW over eight sub-regions under (top to bottom) Historical, RCP2.6, RCP8.5, SSP1-2.6, and SSP5-8.5 scenarios (Unit: m). The period, 1981-2005 was used as reference period for calculating climatology.

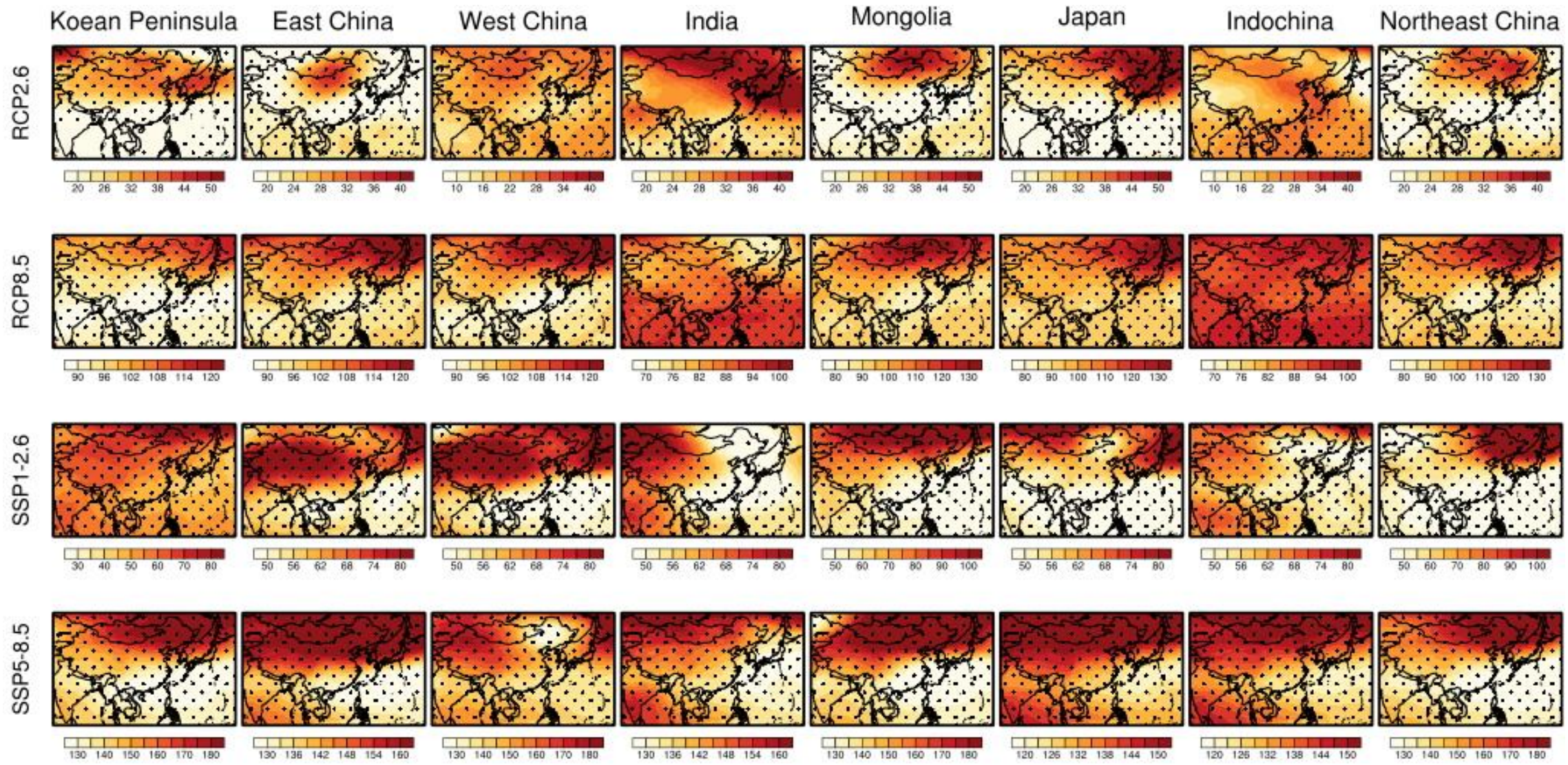

**Figure S10.** Composites of geopotential height at 500hPa anomalies during nighttime for CDNHW over eight sub-regions under (top to bottom) Historical, RCP2.6, RCP8.5, SSP1-2.6, and SSP5-8.5 scenarios (Unit: m). The period, 1981-2005 was used as reference period for calculating climatology.